\documentclass[preprintnumbers,
amsmath,amssymb,floatfix,10pt,prd,twocolumn,
superscriptaddress,longbibliography,nofootinbib]{revtex4-2}
\usepackage{bm}
\usepackage{amsfonts}
\usepackage{latexsym}
\usepackage{amsmath}
\usepackage{palatino}
\usepackage{textcomp}
\usepackage{float}
\usepackage{booktabs}
\usepackage{dcolumn}
\usepackage{ragged2e}
\usepackage{epsfig}
\usepackage[x11names]{xcolor}
\usepackage{hyperref}
\hypersetup{           
    colorlinks=true,                
    breaklinks=true,                
    urlcolor= blue,                
    linkcolor= blue,                
    bookmarksopen=false,
    filecolor=black,
    citecolor=red,
    linkbordercolor=blue}
\usepackage{graphicx}
\usepackage{orcidlink}
\usepackage[T1]{fontenc} % For better handling of hyphenation and accented characters

\begin{document}
%\title{Frolov Black Hole Surrounded by Quintessence - II: Quasinormal modes, Weak Lensing and Chaos}
\title{Frolov Black Hole Surrounded by Quintessence - II: Quasinormal Modes, Greybody Factors, Deflection Angle and Chaos Bound}
\author{Mrinnoy M. Gohain\orcidlink{0000-0002-1097-2124}}
\email{mrinmoygohain19@gmail.com}
\affiliation{%
 Department of Physics, Dibrugarh University, Dibrugarh \\
 Assam, India, 786004}
\affiliation{%
 Department of Physics, DHSK College, Dibrugarh \\
 Assam, India, 786001}
 
\author{Kalyan Bhuyan\orcidlink{0000-0002-8896-7691}}%
 \email{kalyanbhuyan@dibru.ac.in}
\affiliation{%
 Department of Physics, Dibrugarh University, Dibrugarh \\
 Assam, India, 786004}%
 \affiliation{Theoretical Physics Divison, Centre for Atmospheric Studies, Dibrugarh University, Dibrugarh, Assam, India 786004}

\author{Hari Prasad Saikia \orcidlink{0009-0008-9048-7719}}%
 \email{hariprasadsaikia@dibru.ac.in}
\affiliation{%
 Department of Physics, Dibrugarh University, Dibrugarh \\
 Assam, India, 786004}

%\author{Rajnandini Borgohain \orcidlink{0000-0002-4465-7974}}%
% \email{rajnandiniborgohainzr@gmail.com}
%\affiliation{%
% Department of Physics, Dibrugarh University, Dibrugarh \\
% Assam, India, 786004}
% 
%\author{Tonmoyee Gogoi \orcidlink{0000-0002-4465-7974}}%
% \email{tonmoyeegogoi2062@gmail.com}
%\affiliation{%
% Department of Physics, Dibrugarh University, Dibrugarh \\
% Assam, India, 786004}%
% 
%\author{Kakoli Bhuyan \orcidlink{0000-0002-4465-7974}}%
% \email{bhuyankakoli04@gmail.com}
%\affiliation{%
% Department of Physics, Dibrugarh University, Dibrugarh \\
% Assam, India, 786004}%
%
%\author{Prabwal Phukon \orcidlink{0000-0002-4465-7974}}%
% \email{prabwal@gmail.com}
%\affiliation{%
% Department of Physics, Dibrugarh University, Dibrugarh \\
% Assam, India, 786004}%
% \affiliation{Theoretical Physics Divison, Centre for Atmospheric Studies, Dibrugarh University, Dibrugarh, Assam, India 786004}

\keywords{Black Hole; Frolov Black Hole; Black Hole Thermodynamics; Null-Geodesics}
\begin{abstract}
In this work, we have examined the scalar quasinormal modes, greybody factors, weak gravitational lensing, and the chaos bound of a Frolov black hole surrounded by a quintessence field. We studied the quasinormal frequencies and deduced that an increase in the value of the quintessence parameter $c$ dissipates the oscillatory behavior and thus slow down the decay of perturbations. The greybody factors are highly dependent on the multipole moment $l$ and BH parameters, with a larger $q$ hindering the transmission and smaller $\alpha_0$ making the barrier steeper. The deflection angle of weak lensing under thin lens approximation, monotonically decreases as the value of impact parameter $b$ increases, and is visibly affected by the parameter $c$, and that $\alpha_0$ and $q$ has very minimal effect. Lastly, we investigated the chaos bound behaviour of photon trajectories in the near-horizon neighbourhood, which indicates that it is satisfied for bigger circular photon orbits but broken for smaller ones, with high-angular-momentum photons being chaotic. This violation of the bound points towards the role of quintessence in modifying the thermodynamic features of the BH, with possible adjustments to the standard Bekenstein-Hawking entropy.
\end{abstract}

\maketitle
%\newpage
%\tableofcontents
\section{Introduction}
\label{intro}
In theoretical and astrophysical contexts, black holes (BHs) are among the most captivating topics that have gained significant attention in recent decades. After the first direct observation of the BH shadow in the very centre of the M87 galaxy \cite{EventHorizonTelescopeCollaboration2022May} by the Event Horizon Telescope (EHT) collaboration, the importance of BH physics has increased dramatically in recent years. From a historical standpoint, Karl Schwarzschild provided the first known solutions to the Einstein field equations. These solutions, which are later recognised as the Schwarzschild BH solution, indicated the feasibility of a spacetime that contains a spherically symmetric, non-rotating, and chargeless body. The Riessner-Nordstr\"{o}m BH is another generalisation of the uncharged Schwarzschild BH to a charged one. Other typical BH types are the Kerr-Newman BH, which represents a charged and rotating axisymmetric BH solution, and the Kerr BH, which describes a rotating and uncharged BH solution. Some of the other important intriguing physical characteristics of BHs include quasinormal modes, thermodynamics, and optical characteristics like lensing and shadows.

In both classical and quantum backgrounds, the theory of GR is known to be UV incomplete, meaning that singularities are present. The centre of the well-known BH systems, such as Kerr, Riessner Nordstrom, and Schwarzschild, features curvature singularities. This suggested that if one wants a UV complete theory, Einstein's GR must be modified. Numerous attempts have been made to develop alternative versions of a UV complete theory, but each modification has its own set of issues. For example, adding derivative and higher order curvature terms to the gravitational action results in what are known as ``ghost instabilities," where ``ghosts" usually refer to unphysical degrees of freedom. In a seminal study published in 2016 \cite{Frolov2016Nov}, V.L. Frolov developed metrics using a number of intuitive assumptions to determine the existence of non-singular BH solutions without changing the theory of GR. By generalising to the charged case, Frolov presented a number of options in his original study, including the modified Hayward solution. The length scale parameter $\alpha_0$ (denotes by $l$ in his original paper) is related to the critical energy scale $\mu$ since $\alpha_0 = \mu^{-1}$. Thus, in addition to the BH's mass, there is another parameter in his own work called $\alpha_0$ that essentially sets the scale at which the modification of Einstein equations becomes relevant. Technically speaking, at a scale where $\alpha_0^{-2}$ is equivalent to the curvature scalar $R$. The second argument is that the standard metric tensor $g_{\mu \nu}$ can be used, and that there is a length scale $\lambda$ at which quantum gravity effects become prominent. The magnitude of length scale is significantly lower than the parameter $\alpha_0$. Because of these presumptions, the solution is not singular at $r =0$. Song et al. \cite{Song2024Aug} recently, examined the quasinormal modes of a Frolov BH under scalar perturbations, and they found that the QNMs were stable with temporal decay behaviour, where by quantum gravity phenomena takes an important part. Additionally, by constraining the BH parameters in relation to the M87* shadow observational data, Kumar et al. examined the shadow and lensing characteristics of a Frolov BH \cite{Kumar2019Dec}.

In the context of BH physics, BHs are often surrounded by matter, such as accretion disks or possibly exotic matter and fields, which can perturb the surrounding spacetime geometry. These resulting perturbations can change the geodesic motion within the BH spacetime; in other words, the geodesic motion contain the information about the nature of these perturbations. Such perturbations, in turn, will lead the BHs to radiate gravitational waves. It is possible to divide  loosely into three stages the dynamical evolution, namely inspiral, merger and then followed by ringdown; and among them, the most interesting is the damping oscillations of black holes leading to QNMs. They depend on both the intrinsic parameters—the mass, charge, and angular momentum of the BH and the nature of perturbations, for example, the scalar or electromagnetic field surrounding it. The frequencies of QNMs are complex-valued, and the real part gives the oscillation frequency of the perturbation, while the imaginary part gives the damping rate. Since QNMs are intrinsically linked to the black hole parameters, they are a vital tool for the investigation of the basic properties of black holes.

Whether these perturbations grow or decay will determine the stability of the BH. The perturbations in a black hole spacetime result in oscillations with characteristic frequencies, now called quasi-normal frequencies (QNF). These oscillations are in general termed as quasinormal modes because their growth or decay is controlled by the stability of the black hole. QNMs are of interest in a number of contexts. Within the context of the AdS/CFT correspondence \cite{Maldacena1999Apr}, QNMs allow one to study both equilibrium and non-equilibrium properties of strongly coupled thermal gauge theories by using their gravity dual \cite{Son2007Nov}. More precisely, the QNM spectrum of the dual gravitational background corresponds to the poles of retarded correlators in the associated field theory \cite{Nunez2003Jun,Birmingham2002Mar}. QNMs are profoundly important in the understanding of the behaviour of astrophysical BHs and gravitational wave (GW) astronomy. The foundational work on the theory of BH perturbation was first established by Regge and Wheeler \cite{Regge1957Nov}, who showed that the equations of motion of a perturbed BH can be expressed in the form of Schr\"{o}dinger equation of quantum mechanics, which then can be solved through various analytical and numerical techniques. Zerilli \cite{Zerilli1970Nov,
Zerilli1970Mar} extended this framework, and later on Vishveshwara \cite{Vishveshwara1970May} analyzed the Schwarzschild metric stability, identifying the QNM features through GW scattering \cite{Vishveshwara1970Aug}. Chandrasekhar and Detweiler later proved the isospectrality of the Regge-Wheeler and Zerilli potentials and computed Schwarzschild black hole QNMs using the shooting method \cite{Subrahmanyan1975Aug}. Blome and Mashhoon \cite{Blome1984Jan} approximated the effective potential with the Eckart potential to analytically calculate QNMs, while Ferrari and Mashhoon \cite{Ferrari1984Jul} used the P\"{o}schl-Teller potential. Leaver utilized numerical methods for QNM computation by following the Continued Fraction Method (CFM) \cite{Leaver1986May,Leaver1986Jul,LeaverE.1985Dec}. Other approaches, including the Horowitz-Hubeny (HH) method \cite{Horowitz2000Jun} and the Asymptotic Iteration Method (AIM) \cite{Cho2010Jun}, have also been developed. Comprehensive reviews on QNMs and their applications can be found in \cite{Berti2009Jul,Kokkotas1999Sep,Konoplya2011Jul}.
In modified gravity and quantum theories, regular black holes have been proposed \cite{Ayon-Beato1998Jun,Ayon-Beato1999May,Ayon-Beato2000Sep}, often resulting from gravity’s coupling with nonlinear electrodynamics.
QNMs of regular black holes have been analyzed by Flachi and Lemos \cite{Flachi2013Jan}, Toshmatov et al. \cite{Toshmatov2015Apr,Toshmatov2018Oct} and others \cite{Lopez2020Apr,Li2017Jan,Wu2018Apr,
reg_BH,Bueno2024Mar}.

Geodesic motion around black holes includes a whole family of closed and open orbits, depending on their radial position \( r \) from the black hole. There are orbits that could be stable or unstable, depending on an effective potential \( V \) that the particles feel. Another way of looking at the geodesic stability is through the Lyapunov exponents.
The Lyapunov exponents are powerful mathematical tools in the analysis of chaotic dynamics, mainly used for nonlinear systems. Similarly, they can be applied to investigate particle trajectory stability. Mathematically, the Lyapunov exponents \( \lambda \) quantify the average rate of convergence or divergence of nearby geodesics or geodesic congruences. In particular, the principal Lyapunov exponent can be expressed in terms of the second derivative of the potential \( V \) evaluated at the extremum of the potential and yields a direct measure of the stability of the orbits. The Lyapunov exponent in BH systems is constrained with an upper bound, known as the ``chaos bound" first derived by Maldacena et al. \cite{Maldacena2016Aug}. This bound is expressed as 
\(
\lambda \leq \frac{2 \pi T}{\hbar},
\), where $T$ is the BH temperature. 
This bound typically indicates the upper limit on the degree of chaos in quantum thermal systems. Maldacena and his team \cite{Maldacena2016Aug}, arrived to this result through the formalism of quantum field theory and shock wave analysis near the BH horizon. 
Some of the works based on the use of Lyapunov exponents and chaos bound in BH spacetimes can be found in \cite{Pretorius2007May,Cornish2003Apr,
Cardoso2008Dec,Giri2021Aug,Lyu2024Sep,
Barzi2024Aug,Saleem2024Apr,Chowdhury2024Sep,Park2024Apr,
Lei2021Aug,Singh2024May,Lei2024Apr,
Li2024Sep,Prihadi2024Jul,Lu2024Jun,Kumara2024Jan}

The paper is planned as follows: In the first part, in Section \ref{sec2} we briefly discuss the framework of a Frolov BH surrounded by a quintessence field. In Section \ref{sec3}, we shall discuss the effect of scalar perturbation on the Frolov BH and determine the QNMs and hence investigate the influence of the quintessence and other model parameters on the QNM frequencies. In Section \ref{sec4}, we discuss the transmission probability of Hawking radiation in  terms of greybody factors. In Section \ref{sec5}, we study the behaviour of deflection angle of photon rays in the BH spacetime and how the quintessence affects the trajectories. In Section \ref{sec6}, we study the Lyapunov exponent and chaos bound of null geodesics in the vicinity of the Frolov BH and the effect of quintessence on the chaos bound. Finally, in Section \ref{conc} we conclude the study and discuss the results.

\section{Frolov Black Hole Surrounded by Quintessence}
\label{sec2}
The Frolov BH is the charged extension of the Hayward BH described by the metric \cite{Song2024Aug}
\begin{equation}
ds^2 = - f(r) dt^2 + f(r)^{-1} dr^2 + r^2 d\Omega^2,
\label{line_el}
\end{equation}
where 
\begin{equation}
f(r) = 1 - \frac{(2Mr - q^2)r^2}{r^4 + (2Mr + q^2)\alpha_0^2},
\label{lapse}
\end{equation} and $d\Omega^2 = d\theta^2 + \sin^2 \theta d\phi^2$. Here $M$ is the ADM mass of the BH. The Frolov BH is associated with the cosmological constant $\Lambda = 3/\alpha_0^2$, where $\alpha_0$ is known as the Hubble length (length scale parameter) \cite{Song2024Aug,Hayward2006Jan}, that would serve as a free parameter of the model. The Hubble length display itself as a Universal hair and is restricted by $\alpha_0 \le \sqrt{16/27}M$. In our analysis, we shall set $M = 1$ for the sake of simplicity. Another free parameter included in the Frolov BH is the charge parameter $q$, that satisfies the constraint $0\le q \le 1.$ Clearly, in the limiting value of $q \to 0$, the Frolov BH reduces to the Reissner-Nordstrom(ND) BH solution. Also as $\alpha_0 \to 0$, one obtains the Schwarzschild BH solution. 

Kiselev had suggested an interesting possibility that a quintessence field can affect the properties of the BH. The quintessence field follows the relations \cite{Kiselev2003Mar}
\begin{equation}
T^{\phi}_{\phi} = T^{\theta}_{\theta} = -\frac{1}{2}(3w + 1) T^{r}_{r} = \frac{1}{2}(3w + 1) T^{t}_{t}
\end{equation}
where $w$ indicates the equation of state (EoS) parameter of the quintessence field. In general, the EoS parameter $w$ for quintessence field falls within the range $-1 < w < -1/3$.

In this work, which is a follow up of our previous work \cite{Gohain2024Dec}, we shall study the QNMs and chaos bound scenario. We shall try to investigate whether the aforementioned aspects are influenced by the presence of the quintessence field and if yes, how do the QNM frequencies are affected by the the quintessence field.

Following the procedure of Kiselev, the effective metric in the presence of quintessence can be obtained by incorporating the term $-c/r^{3w+1}$ into the BH metric, Eq. \eqref{lapse} as: 
\begin{equation}
f(r) = 1 - \frac{(2Mr - q^2)r^2}{r^4 + (2Mr + q^2)\alpha_0^2} - \frac{c}{r^{3w + 1}},
\label{lapse_quin}
\end{equation}
In the modified lapse function given by Eq. \eqref{lapse_quin}. Following this same technique, few of the recent works are carried out are \cite{Zeng2020Nov,Mustafa2022Dec,Belhaj2020Oct,Chen2022Apr,Toshmatov2017Feb} (and references therein).  In our work, we shall fix the value of the EoS parameter $w = -2/3$. Here, $c$ is a constant that describes a coupling of the quintessence field with our concerned BH system. Physically, this may represent the intensity of the quintessence field, and we shall consider it as a free parameter and see how it plays a role in the behaviour of the QNMs, greybody bounds, deflection angle and chaotic orbits.

\section{Quasinormal modes of a Frolov BH surrounded by Quintessence}
\label{sec3}
In this section, let us discuss the properties of the QNMs of the Frolov BH and how they are governed by the presence of quintessence field. First of all, let us start by studying how the BH system responds to the perturbations to the massless scalar field $\Phi$. The evolution of the scalar field is governed by the Klein-Gordon (KG) equation, which is
\begin{equation}
\nabla_\mu \nabla^\mu \Phi = \frac{1}{\sqrt{-g}} \partial_\nu \left( g^{\mu \nu} \sqrt{-g} \partial_\mu \Phi \right) = 0,
\label{KG eqn}
\end{equation}
Since the spacetime is spherically symmetric, we may use the method of separation of variables involving spherical harmonics $Y_{l,m} (\theta, \phi)$, with $l$ and $m$ denoting the angular and azimuthal quantum numbers repectively. Then, the solution for $\Phi$ takes the form \cite{Fernando2012Sep,Jusufi2020Apr}
\begin{equation}
\Phi (t, r, \theta, \phi) = \sum_{l,m} \exp (-i\omega t) Y_{l,m}(\theta, \phi) \frac{R(r)}{r}.
\label{soln_KG}
\end{equation}
This solution renders the KG equation \eqref{KG eqn} to the form

\begin{equation}
\frac{1}{r^2} \frac{d}{dr} \left[ r^2 f(r) \frac{d}{dr}\left(\frac{R(r)}{r}\right) \right] + \left( \frac{\omega^2}{f(r)} - \frac{l(l+1)}{r^2} \right) \frac{R(r)}{r} = 0.
\label{schrod1}
\end{equation}
The above Eq. \eqref{schrod1} can be restructured into the familiar Schr\"{o}dinger like form:
\begin{equation}
\frac{d^2 R(\ast)}{d r_{\ast}^2} + \left(\omega^2 - V(r_\ast)\right) R(r_\ast) = 0,
\label{schrod2}
\end{equation}
where $V(r_\ast)$ is the effective potential given by:
\begin{equation}
V(r_\ast) = f(r) \frac{l(l+1)}{r^2} + \frac{f(r)f'(r)}{r},
\label{eff_pot1}
\end{equation}
Here, the angular quantum number $l$ can take values 0, 1, 2,\dots and $r_{\ast}$ denotes the tortoise coordinate, which is related to radial coordinate $r$ through $dr_{\ast}/dr = 1/f(r)$. It spans from $-\infty$ at the event horizon to
$+ \infty$ towards spatial infinity. It is customary to adopt the simplified notation $r_\ast \to  x, \, R(r_\ast) \to \psi(x)$ and $\omega^2 - V(r_\ast) \to Q(x)$. The potential $Q(x)$ takes constant values at $x =\pm \infty$, but not necessarily symmetric at both the ends, and attains a maximum value at some point $x = x_0$.

%Now, to determine the QNMs, the boundary conditions are: 
%\begin{equation}
%\Psi(r_\ast) \propto e^{\pm i \omega r_\ast}, \quad r_\ast \to \pm \infty.
%\label{boundaryconditions}
%\end{equation}
%The plus and minus sign represents to the outgoing and ingoing waves at the boundary respectively. There are numerous ways to calculate the QNMs like Wentzel-Kramers-Brillouin (WKB) approximation, Leaver's method of continued fraction, asymptotic iteration method (AIM), pseudo-spectral (PS) method, Horowitz-Hubeny method, Mashoon method etc. Out of all these methods, the WKB method remains a precise and efficient approach in terms of least errors. Recent works have developed the method from 1st upto 13th order. However, the higher order does not automatically ensure best accuracy. For example, [zhang] has found that the 9$^{th}$ order WKB exhibit least error. Despite being highly accurate, the WKB method loses its effectiveness for higher overtones (denoted by $n$), especially when $n > l$. In this work, we shall deal with the determination of quasinormal frequencies using WKB approximation of 6$^{th}$ and 9$^{th}$ order and then perform a comparative analysis between the two.  While doing the comparative analysis, it is useful to estimate the relative error between the two methods which is given as:
%\begin{equation}
%\varepsilon_{WKB} = \left|\frac{\omega_{WKB(9^{th})} - \omega_{WKB(6^{th})}}{\omega_{WKB(9^{th})}} \right| \times 100 \%. 
%\end{equation}

The Eq. \eqref{schrod2} has an exact solution when  $Q(x)$ is constant. If $Q(x)$ relies on $x$, then the equation is exactly solvable for some particular forms of $Q(x)$, and hence for $V(r_\ast)$. In [83], the WKB approximation was used to solve the Schrödinger-like problem \eqref{schrod2}. The WKB approach is often applicable when the effective potential varies slowly [84].

Conventionally, the formulation of the WKB problem involves obtaining the solutions for incident, reflected, and transmitted amplitudes, with incident and reflected amplitudes being of equal size. In the context of black hole perturbation, the incident amplitude is zero. As a result, reflected and transmitted amplitudes are found to be of the same order of magnitude. This follows that the WKB approximation is applicable given the following condition:
\begin{equation}
\frac{d Q(x)}{dx} \Bigg|_{x = x_0} = 0.
\label{cond_wkb}
\end{equation}
If the the turning points are too close, the WKB approximation will become inapplicable. The only possible technique is to match the solutions across each turning point simultaneously. Moreover, the potential in between the two turning point is approximated to a parabolic shape \cite{Schutz1985Apr}. The matching of the solutions in the different regions is done by employing Taylor expansion of the potential about its maximum, \( x_0 \), as can be easily found in any quantum mechanics textbooks \cite{Zettili2022Sep,Griffiths2018Aug}.

The solution in the region between the turning points can be obtained by expanding \( Q(x) \) around the maximum \( x_0 \). As the solution must be continuous, it is possible to obtain the approximate solutions for the two asymptotic regions where \( Q(x) \to \omega \), a constant, at large distances. The matching conditions gives what is usually known as the quasinormal mode (QNM) condition, because it leads to discrete and complex frequencies called quasinormal frequencies. The condition is given by \cite{Iyer1987Jun,Iyer1987Jun1}
\begin{equation}
\frac{Q(x_0)}{\sqrt{2Q''(x_0)}} = - i \left(n + \frac{1}{2}\right)
\label{condqnm}
\end{equation}
Now if one uses the condition \eqref{condqnm} in Eq. \eqref{schrod2}, it leads to
\begin{equation}
\frac{\omega^2 - V(r_0)}{\sqrt{-2V''(r_0)}} = -i \left(n+\frac{1}{2}\right),
\label{omeg1}
\end{equation}
or
\begin{equation}
\omega^2 = V(r_0) - i \left( n + \frac{1}{2}\right)\sqrt{-2V''(r_0)} \equiv X - i Y.
\label{omeg2}
\end{equation}
where $r_0$ represents the maxima of the effective potential. Here, we have also designated $X = V(r_0)$ and $Y = \left( n + \frac{1}{2}\right)\sqrt{-2V''(r_0)}$. This leads to the complex quasinormal frequencies given by
\begin{equation}
\begin{aligned}
&\omega = \omega_R - i \omega_I; \\& \text{ with } \quad  \omega_R = \frac{\sqrt{\sqrt{X^2+Y^2}+X}}{\sqrt{2}}, \, \\& \text{ and }  \omega_I = \frac{Y}{\sqrt{2} \sqrt{\sqrt{X^2+Y^2}+X}}
\end{aligned}
\label{omeg_mode}
\end{equation}
In our work, we follow the first-order WKB approximation. It is also possible to calculate the higher order WKB approximates, which is currently refrained from this analysis for simplicity. Higher order WKB techniques can be found in refs \cite{Konoplya2003Jul,Konoplya2023Mar,
Konoplya2019Jul}. 
\begin{figure*}[htb]
\centerline{\includegraphics[scale=0.45]{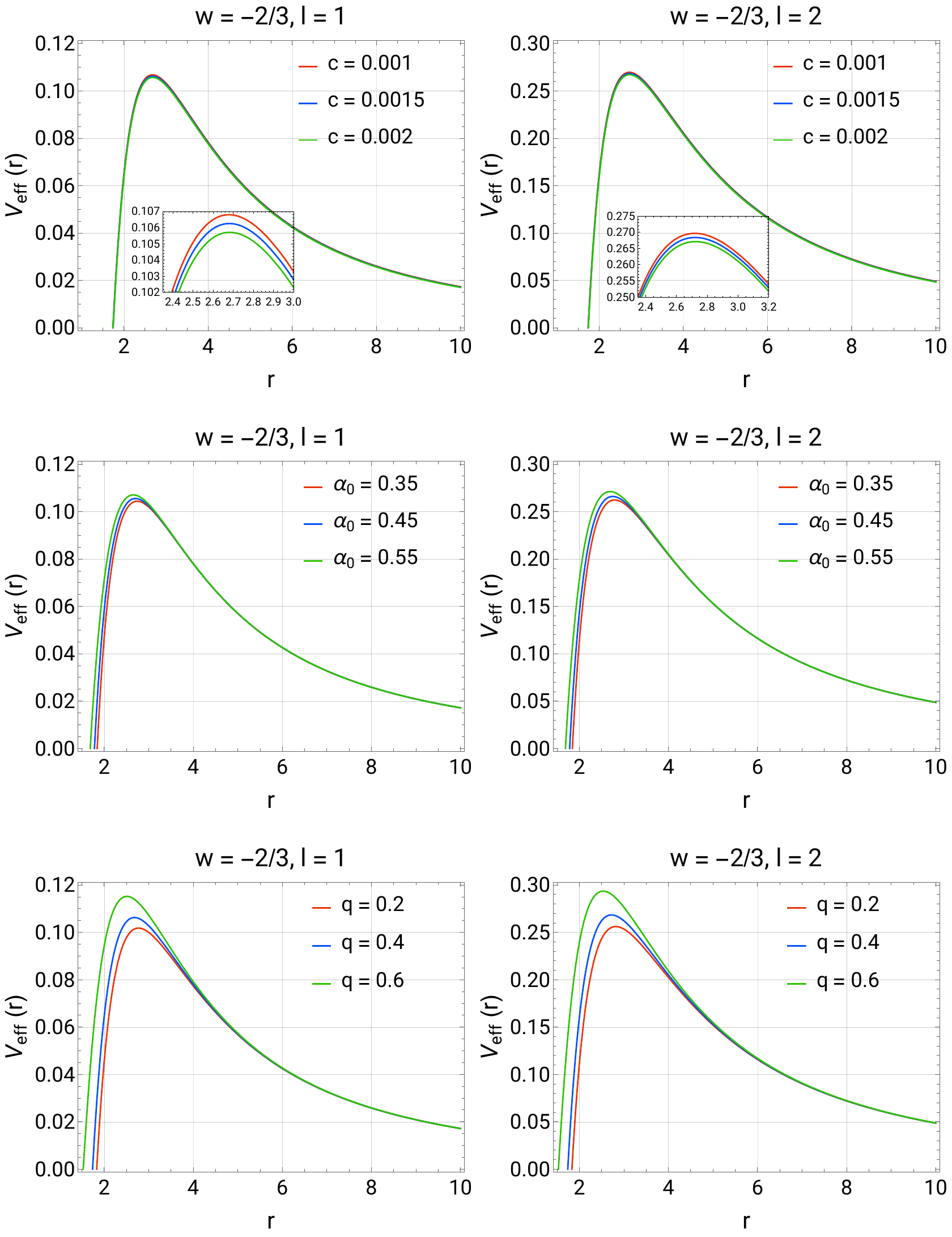}}
\caption{The plot of the effective potential is shown for different combinations of the parameters $c$, $\alpha_0$ and $q$ for $l = 1$ and $2$. In the top panel, $\alpha_0$ and $q$ are set fixed at $0.5$ and $0.4$ respectively. In the middle panel, $c$ and $q$ are set at $0.015$ and $0.4$ and in the bottom panel, $c$ and $\alpha_0$ are fixed at $0.015$ and $0.5$ respectively.}
\label{eff_pot_all}
\end{figure*}
%\begin{figure*}[htb]
%\centerline{\includegraphics[scale=0.40]{eff_pot_alp.pdf}}
%\caption{Effective potential 2}
%\label{eff_alp}
%\end{figure*}
%\begin{figure*}[htb]
%\centerline{\includegraphics[scale=0.40]{eff_pot_q.pdf}}
%\caption{Effective potential 3}
%\label{eff_q}
%\end{figure*}

 In the Fig. \ref{eff_pot_all}, we present the radial evolution of the effective potential \(V_{\text{eff}}\) for multipole moments \(l = 1\) and \(l = 2\). Here, as previously stated, we have fixed an EoS parameter \(w = -2/3\). We observe the impact of the parameters \(c\), \(\alpha_0\), and \(q\) from the figure as follows:

\begin{itemize}
\item \textbf{Quintessence parameter \(c\):} With the increase in the value of the quintessence parameter \(c\), the height of the peak of \(V_{\text{eff}}\) declines for the case \(l = 1\) and for \(l = 2\) the change in the height of the peak is moderate.
\item \textbf{Length Scale Parameter \(\alpha_0\):} The effective potential is seen to be quite insensitive to \(\alpha_0\) values for \(l = 1\); but nonetheless, a slight variation in the position of the peak towards increasing values of \(r\) is noticed as \(\alpha_0\) decreases. This behaviour is also noticed in the \(l = 2\) case, where the height of the peak increases slowly with increasing \(\alpha_0\).
\item \textbf{Charge Parameter \(q\):} The peak value of \(V_{\text{eff}}\) in this case turns out to be highly sensitive to the change in values of \(q\) for both \(l = 1\) and \(l = 2\). For both situations, as \(q\) decreases, it reduces the peak value along with shifting the position of the peak towards increasing \(r\) values.
\end{itemize}

%\begin{table*}[ht]
%\centering
%\begin{tabular}{||c | c | c ||}
%\hline
%\textbf{$c$} & \textbf{$\omega (w = -4/9)$} & \textbf{$\omega (w = -2/3)$} \\
%\hline
%$0.001$ & $0.587217 - 0.271786i$ & $0.586202 - 0.271946i$ \\
%$0.002$ & $0.585979 - 0.271448i$ & $0.583947 - 0.271768i$ \\
%$0.003$ & $0.584741 - 0.271108i$ & $0.581687 - 0.271588i$ \\
%$0.004$ & $0.583504 - 0.270769i$ & $0.579423 - 0.271408i$ \\
%$0.005$ & $0.582267 - 0.270429i$ & $0.577156 - 0.271227i$ \\
%$0.006$ & $0.581029 - 0.270089i$ & $0.574884 - 0.271045i$ \\
%$0.007$ & $0.579792 - 0.269748i$ & $0.572609 - 0.270863i$ \\
%$0.008$ & $0.578556 - 0.269407i$ & $0.570329 - 0.270679i$ \\
%$0.009$ & $0.577319 - 0.269065i$ & $0.568046 - 0.270495i$ \\
%$0.010$ & $0.576082 - 0.268723i$ & $0.565758 - 0.270310i$ \\
%$0.011$ & $0.574846 - 0.268381i$ & $0.563466 - 0.270124i$ \\
%$0.012$ & $0.573610 - 0.268038i$ & $0.561170 - 0.269938i$ \\
%$0.013$ & $0.572373 - 0.267694i$ & $0.558869 - 0.269750i$ \\
%$0.014$ & $0.571137 - 0.267350i$ & $0.556565 - 0.269562i$ \\
%$0.015$ & $0.569901 - 0.267006i$ & $0.554256 - 0.269373i$ \\
%$0.016$ & $0.568666 - 0.266662i$ & $0.551942 - 0.269183i$ \\
%$0.017$ & $0.567430 - 0.266317i$ & $0.549624 - 0.268993i$ \\
%$0.018$ & $0.566194 - 0.265971i$ & $0.547302 - 0.268801i$ \\
%$0.019$ & $0.564959 - 0.265625i$ & $0.544975 - 0.268609i$ \\
%$0.020$ & $0.563724 - 0.265279i$ & $0.542644 - 0.268416i$ \\
%\hline
%\end{tabular}
%\caption{Quasinormal mode frequencies $\omega$ for different $c$ and $w$ for $l = 2$.}
%\label{tab_qnm2}
%\end{table*}

%\clearpage
%table for varying c, l =1

\begin{table*}[ht]
\centerline{
\begin{tabular}{||c | c ||}
\hline
\textbf{$c$} & \textbf{$\omega (w = -2/3)$} \\
\hline
$0.001$ & $0.415281 - 0.256251i$ \\
$0.002$ & $0.413811 - 0.255994i$ \\
$0.003$ & $0.412338 - 0.255735i$ \\
$0.004$ & $0.410863 - 0.255473i$ \\
$0.005$ & $0.409386 - 0.255209i$ \\
$0.006$ & $0.407907 - 0.254943i$ \\
$0.007$ & $0.406425 - 0.254674i$ \\
$0.008$ & $0.404941 - 0.254403i$ \\
$0.009$ & $0.403455 - 0.254130i$ \\
$0.010$ & $0.401966 - 0.253854i$ \\
$0.011$ & $0.400475 - 0.253576i$ \\
$0.012$ & $0.398982 - 0.253295i$ \\
$0.013$ & $0.397486 - 0.253011i$ \\
$0.014$ & $0.395987 - 0.252725i$ \\
$0.015$ & $0.394486 - 0.252437i$ \\
$0.016$ & $0.392982 - 0.252146i$ \\
$0.017$ & $0.391475 - 0.251852i$ \\
$0.018$ & $0.389966 - 0.251555i$ \\
$0.019$ & $0.388454 - 0.251256i$ \\
$0.020$ & $0.386940 - 0.250954i$ \\
\hline
\end{tabular} \hspace{0.5cm}
\begin{tabular}{||c | c ||}
\hline
\textbf{$c$} & \textbf{$\omega (w = -2/3)$} \\
\hline
$0.001$ & $0.586202 - 0.271946i$ \\
$0.002$ & $0.583947 - 0.271768i$ \\
$0.003$ & $0.581687 - 0.271588i$ \\
$0.004$ & $0.579423 - 0.271408i$ \\
$0.005$ & $0.577156 - 0.271227i$ \\
$0.006$ & $0.574884 - 0.271045i$ \\
$0.007$ & $0.572609 - 0.270863i$ \\
$0.008$ & $0.570329 - 0.270679i$ \\
$0.009$ & $0.568046 - 0.270495i$ \\
$0.010$ & $0.565758 - 0.270310i$ \\
$0.011$ & $0.563466 - 0.270124i$ \\
$0.012$ & $0.561170 - 0.269938i$ \\
$0.013$ & $0.558869 - 0.269750i$ \\
$0.014$ & $0.556565 - 0.269562i$ \\
$0.015$ & $0.554256 - 0.269373i$ \\
$0.016$ & $0.551942 - 0.269183i$ \\
$0.017$ & $0.549624 - 0.268993i$ \\
$0.018$ & $0.547302 - 0.268801i$ \\
$0.019$ & $0.544975 - 0.268609i$ \\
$0.020$ & $0.542644 - 0.268416i$ \\
\hline
\end{tabular}}
\caption{Quasinormal mode frequencies $\omega$ for different $c$ for $l = 1$ (left) and $l = 2$ (right).}
\label{tab_qnm_c}
\end{table*}

%for alp
%table for varying alpha0

\begin{table*}[ht]
\centerline{
\begin{tabular}{||c | c ||}
\hline
\textbf{$\alpha_0$} & \textbf{$\omega (w = -2/3)$} \\
\hline
$0.010000$ & $0.412891 - 0.260153i$ \\
$0.038421$ & $0.412902 - 0.260136i$ \\
$0.066842$ & $0.412924 - 0.260101i$ \\
$0.095263$ & $0.412959 - 0.260045i$ \\
$0.123684$ & $0.413006 - 0.259969i$ \\
$0.152105$ & $0.413065 - 0.259872i$ \\
$0.180526$ & $0.413136 - 0.259753i$ \\
$0.208947$ & $0.413217 - 0.259611i$ \\
$0.237368$ & $0.413310 - 0.259443i$ \\
$0.265789$ & $0.413414 - 0.259247i$ \\
$0.294211$ & $0.413528 - 0.259022i$ \\
$0.322632$ & $0.413651 - 0.258765i$ \\
$0.351053$ & $0.413782 - 0.258471i$ \\
$0.379474$ & $0.413921 - 0.258136i$ \\
$0.407895$ & $0.414065 - 0.257756i$ \\
$0.436316$ & $0.414213 - 0.257324i$ \\
$0.464737$ & $0.414363 - 0.256832i$ \\
$0.493158$ & $0.414511 - 0.256270i$ \\
$0.521579$ & $0.414653 - 0.255627i$ \\
$0.550000$ & $0.414784 - 0.254887i$ \\
\hline
\end{tabular} \hspace{0.5cm} 
\begin{tabular}{||c | c ||}
\hline
\textbf{\(\alpha_0\)} & \textbf{\(\omega (w = -2/3)\)} \\ 
\hline
0.010000  & 0.576357 - 0.273981i \\ 
0.038421  & 0.576402 - 0.273976i \\ 
0.066842  & 0.576501 - 0.273964i \\ 
0.095263  & 0.576654 - 0.273946i \\ 
0.123684  & 0.576861 - 0.273921i \\ 
0.152105  & 0.577122 - 0.273887i \\ 
0.180526  & 0.577438 - 0.273844i \\ 
0.208947  & 0.577810 - 0.273791i \\ 
0.237368  & 0.578239 - 0.273724i \\ 
0.265789  & 0.578725 - 0.273644i \\ 
0.294211  & 0.579270 - 0.273545i \\ 
0.322632  & 0.579874 - 0.273426i \\ 
0.351053  & 0.580539 - 0.273282i \\ 
0.379474  & 0.581267 - 0.273108i \\ 
0.407895  & 0.582058 - 0.272897i \\ 
0.436316  & 0.582914 - 0.272643i \\ 
0.464737  & 0.583837 - 0.272334i \\ 
0.493158  & 0.584827 - 0.271959i \\ 
0.521579  & 0.585885 - 0.271501i \\ 
0.550000  & 0.587011 - 0.270940i \\ 
\hline
\end{tabular}}
\caption{Quasinormal mode frequencies $\omega$ for different $\alpha_0$ for $l = 1$ (left) and $l = 2$ (right).}
\label{tab_qnm_alpha}
\end{table*}

%for q
%table for varying q

\begin{table*}[ht]
\centerline{
\begin{tabular}{||c | c ||}
\hline
\textbf{\(q\)} & \textbf{\(\omega (w = -2/3)\)} \\ 
\hline
0.010000  & 0.412891 - 0.260153i \\ 
0.038421  & 0.412902 - 0.260136i \\ 
0.066842  & 0.412924 - 0.260101i \\ 
0.095263  & 0.412959 - 0.260045i \\ 
0.123684  & 0.413006 - 0.259969i \\ 
0.152105  & 0.413065 - 0.259872i \\ 
0.180526  & 0.413136 - 0.259753i \\ 
0.208947  & 0.413217 - 0.259611i \\ 
0.237368  & 0.413310 - 0.259443i \\ 
0.265789  & 0.413414 - 0.259247i \\ 
0.294211  & 0.413528 - 0.259022i \\ 
0.322632  & 0.413651 - 0.258765i \\ 
0.351053  & 0.413782 - 0.258471i \\ 
0.379474  & 0.413921 - 0.258136i \\ 
0.407895  & 0.414065 - 0.257756i \\ 
0.436316  & 0.414213 - 0.257324i \\ 
0.464737  & 0.414363 - 0.256832i \\ 
0.493158  & 0.414511 - 0.256270i \\ 
0.521579  & 0.414653 - 0.255627i \\ 
0.550000  & 0.414784 - 0.254887i \\ 
\hline
\end{tabular} \hspace{0.5cm} 
\begin{tabular}{||c | c ||}
\hline
\textbf{\(q\)} & \textbf{\(\omega (w = -2/3)\)} \\ 
\hline
0.010000  & 0.576357 - 0.273981i \\ 
0.038421  & 0.576402 - 0.273976i \\ 
0.066842  & 0.576501 - 0.273964i \\ 
0.095263  & 0.576654 - 0.273946i \\ 
0.123684  & 0.576861 - 0.273921i \\ 
0.152105  & 0.577122 - 0.273887i \\ 
0.180526  & 0.577438 - 0.273844i \\ 
0.208947  & 0.577810 - 0.273791i \\ 
0.237368  & 0.578239 - 0.273724i \\ 
0.265789  & 0.578725 - 0.273644i \\ 
0.294211  & 0.579270 - 0.273545i \\ 
0.322632  & 0.579874 - 0.273426i \\ 
0.351053  & 0.580539 - 0.273282i \\ 
0.379474  & 0.581267 - 0.273108i \\ 
0.407895  & 0.582058 - 0.272897i \\ 
0.436316  & 0.582914 - 0.272643i \\ 
0.464737  & 0.583837 - 0.272334i \\ 
0.493158  & 0.584827 - 0.271959i \\ 
0.521579  & 0.585885 - 0.271501i \\ 
0.550000  & 0.587011 - 0.270940i \\ 
\hline
\end{tabular}}
\caption{Quasinormal mode frequencies \(\omega\) for different \(q\) and \(w = -2/3\) for \(l = 1\) (left) and \( l = 2\) (right).}
\label{tab_qnm_q}
\end{table*}

%\clearpage

%\clearpage
%for c

%qnm_plots
\begin{figure*}[htb]
\centerline{\includegraphics[scale=0.55]{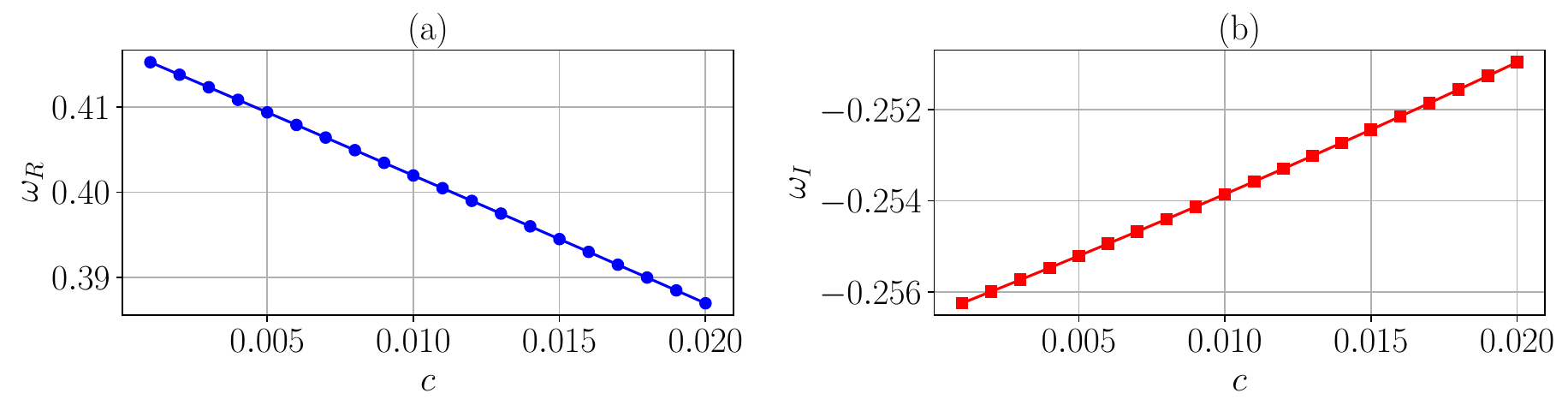}}
\caption{Plot of real and imaginary components for $w = -2/3$ and $n = 0, \, l = 1, \, \alpha_0 = 0.5$ and $q = 0.4$.}
\label{qnm_wkb_c_l1}
\end{figure*}
\begin{figure*}
\centerline{\includegraphics[scale=0.55]{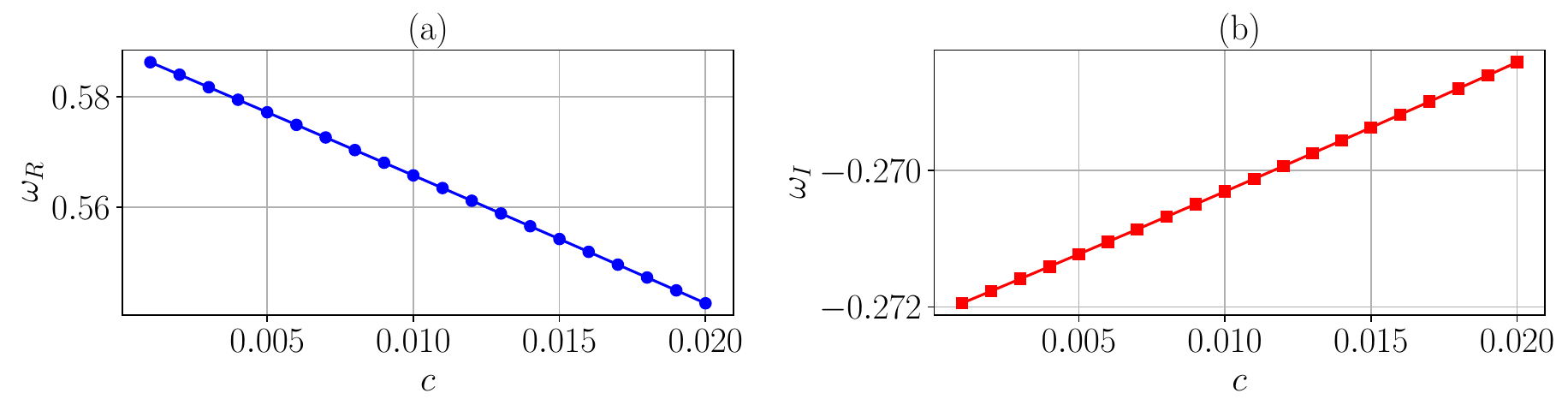}}
\caption{Plot of real and imaginary components for $w = -2/3$ and $n = 0, \, l = 2, \, \alpha_0 = 0.5$ and $q = 0.4$.}
\label{qnm_wkb_c_l2}
\end{figure*}

%for alpha
\begin{figure*}[htb]
\centerline{\includegraphics[scale=0.55]{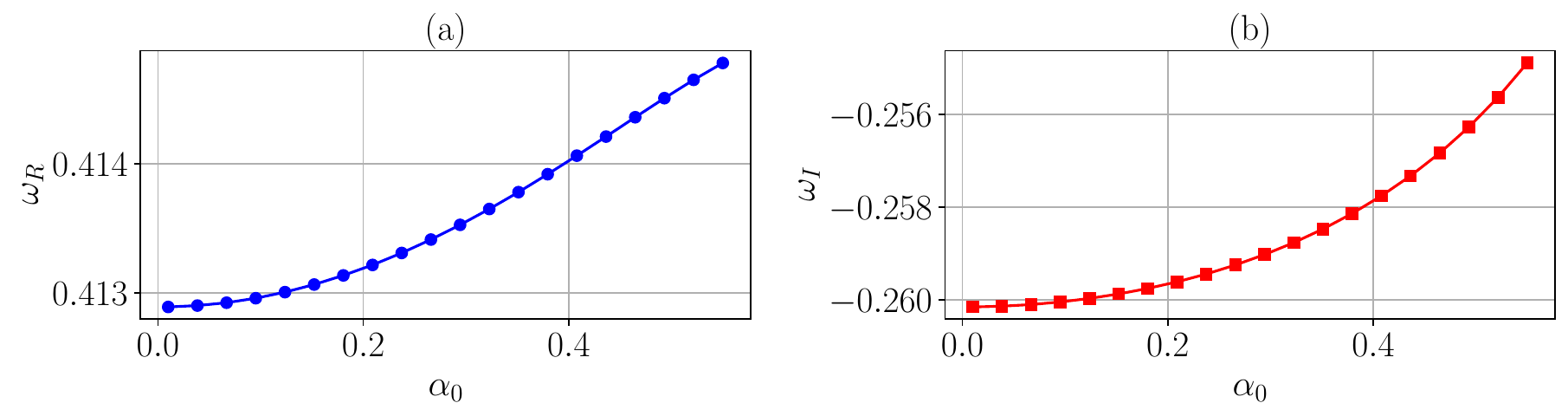}}
\caption{Plot of real and imaginary components for $w = -2/3$ and $n = 0, \, l = 1, \, c = 0.0015$ and $q = 0.4$.}
\label{qnm_wkb_alp_l1}
\end{figure*}
\begin{figure*}
\centerline{\includegraphics[scale=0.55]{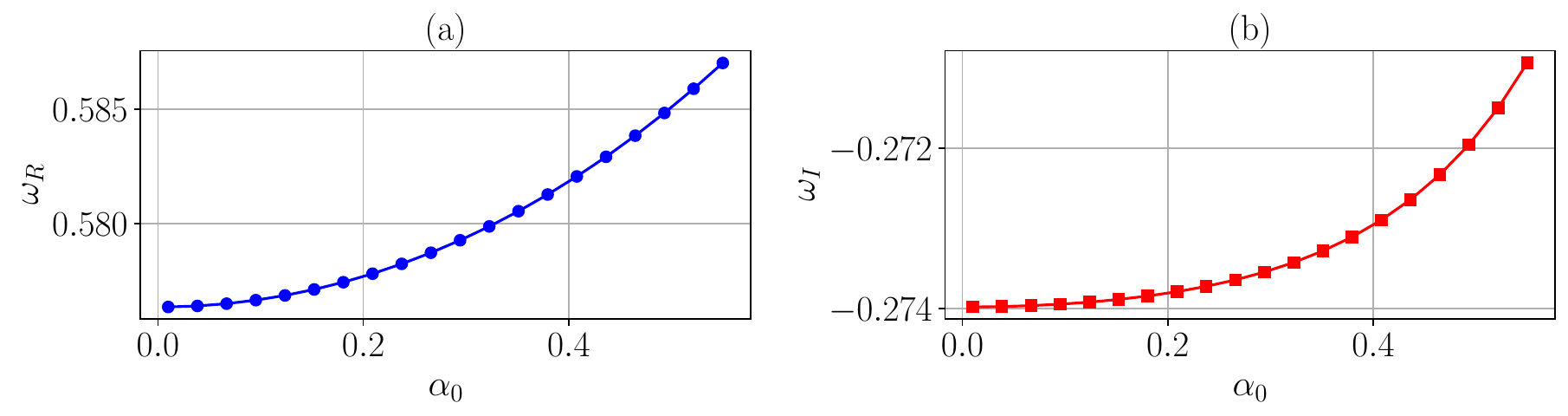}}
\caption{Plot of real and imaginary components for $w = -2/3$ and $n = 0, \, l = 2, \, c = 0.0015$ and $q = 0.4$.}
\label{qnm_wkb_alp_l2}
\end{figure*}

%for q

\begin{figure*}[htb]
\centerline{\includegraphics[scale=0.55]{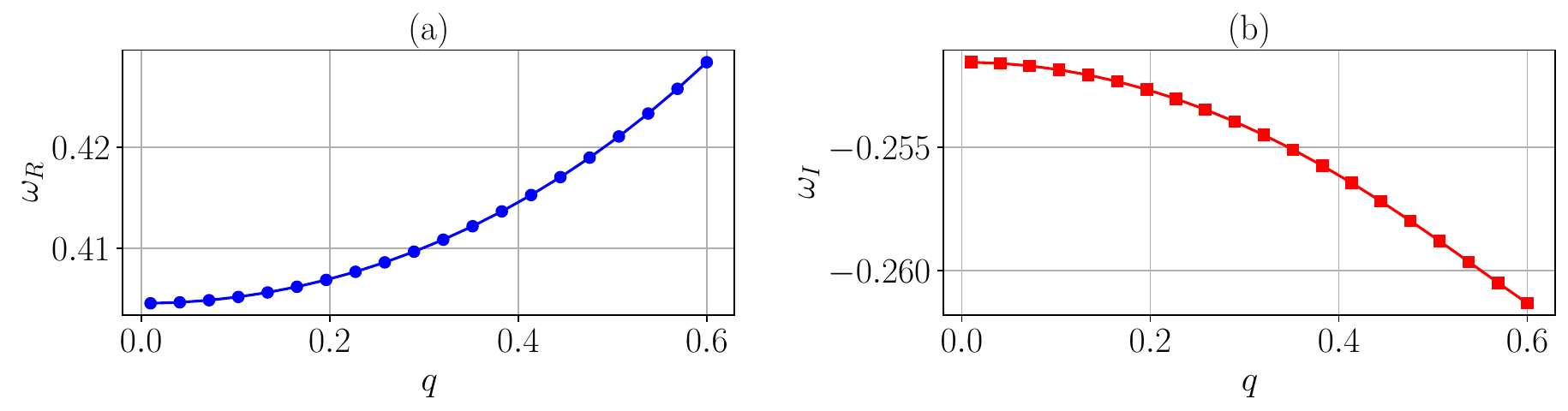}}
\caption{Plot of real and imaginary components for $w = -2/3$ and $n = 0, \, l = 1, \, \alpha_0 = 0.5$ and $c = 0.0015$}
\label{qnm_wkb_q_l1}
\end{figure*}
\begin{figure*}
\centerline{\includegraphics[scale=0.55]{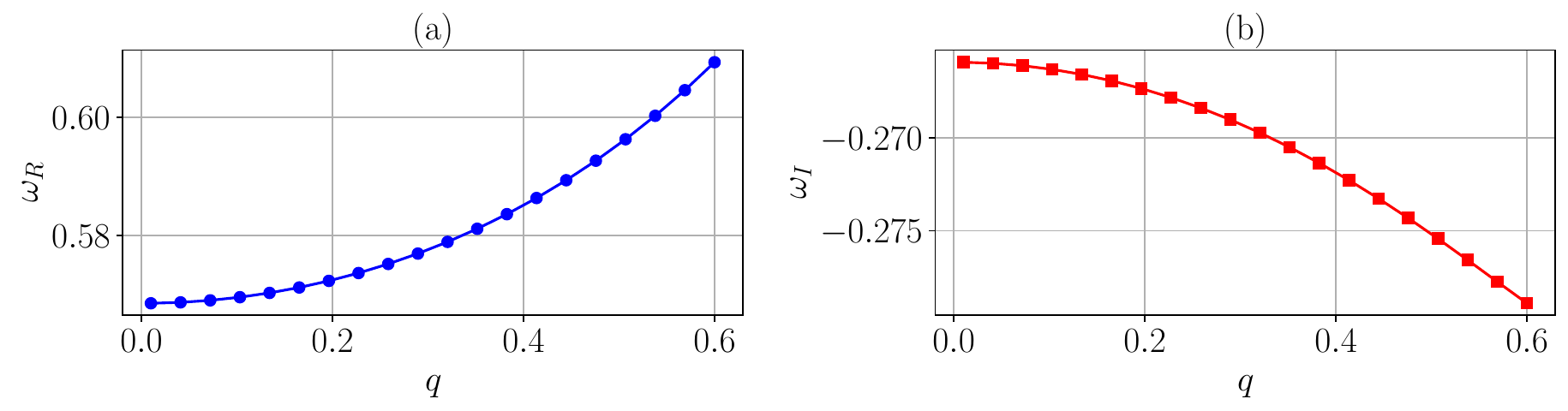}}
\caption{Plot of real and imaginary components for $w = -2/3$ and $n = 0, \, l = 2, \, \alpha_0 = 0.5$ and $c = 0.0015$.}
\label{qnm_wkb_q_l2}
\end{figure*}

The results of the calculated quasinormal modes (QNMs) are presented in Tabs. \ref{tab_qnm_c}, \ref{tab_qnm_alpha} and \ref{tab_qnm_q}, which show the dependence of the QNM frequencies \(\omega\) on the parameter \(c\), \(\alpha\) and \(q\) respectively for a fixed EoS parameter \(w = -\frac{2}{3}\). These are also plotted and shown in Figs. \ref{qnm_wkb_c_l1}, \ref{qnm_wkb_c_l2}, \ref{qnm_wkb_alp_l1}, \ref{qnm_wkb_alp_l2}, \ref{qnm_wkb_q_l1} and \ref{qnm_wkb_q_l2}. We calculated these QNM frequencies using first order WKB technique for the multipole indices \(l = 1\) and \(l = 2\) to illustrate how the spacetime perturbations evolve in the presence of the Frolov BH in the presence of a quintessence field, with \(c\) as the intensity of the field.
The real part of the QNM frequency, \(\text{Re}(\omega)\), that defines the oscillation frequency of the perturbations, is found to decrease monotonically as the value of \(c\) increases. This behaviour means that as the intensity of the quintessence field increases, the oscillation frequency of the perturbations becomes progressively smaller, thereby leading to a weaker response of the BH spacetime.
Whereas in contrast, the imaginary part \(\text{Im}(\omega)\), that is related to the dissipative rate of the oscillations, becomes less negative with increasing values of \(c\). This indicates that a higher intensity of the quintessence field leads to a slower dissipation rate of the perturbations, which implies that the oscillations sustain for a longer period of time. Such a behaviour can be attributed to the enhanced effect of the quintessence field on the effective potential, which in turn affects the propagation and decay properties of the perturbations.

Furthermore, the QNM frequencies as a function of the length scale parameter \(\alpha_0\) and the charge \(q\) are also analyzed. It is observed that \(\text{Re}(\omega)\) increases with increasing values of \(\alpha_0\), which leads to a higher oscillation frequency and thus suggests improved stability of the oscillations for larger values of \(\alpha_0\). In the meanwhile, as \(\text{Im}(\omega)\) becomes more negative, with increasing \(\alpha_0\) values, implies a slower dissipation of the perturbations and more a extended oscillatory behavior. In contrast, the effect of the charge \(q\) on the QNM frequencies is opposite to that of \(\alpha_0\); as \(q\) increases, \(\text{Re}(\omega)\) also increases, signifying that the system oscillates faster. Simultaneously, \(\text{Im}(\omega)\) becomes less negative with higher \(q\), meaning that the perturbations decay more slowly. Loosely speaking, the electromagnetic field coupled to the charge parameter in a way ``tightens" the effective potential around the BH, thus increasing the oscillatory response and prolongs the time of the oscillations.

In general, the monotonic decrease of the real and imaginary parts of \( \omega \) with increasing \( c \) suggests that a strong quintessence field results in smaller oscillation frequencies and longer-lasting perturbations. The combination of \( c \), \( \alpha_0 \), and \( q \)  determines the stability and dissipative behaviour of the BH perturbations, which justifies the usefulness of QNM frequencies as sensitive methods for investigating the role of electromagnetic interactions in the physics of BH spacetimes.

\section{Greybody Factors}
\label{sec4}
Hawking proved that a BH can emit thermal radiation at its event horizon such that it is nearly a blackbody spectrum, when quantum effects are taken into consideration. This radiation was later termed as Hawking radiation \cite{Hawking1975Aug} which interacts with the spacetime curvature around the BH as it departs from the BH, thereby modifying its properties. So, the usual blackbody spectrum of the Hawking radiation cannot be detected by an observer located at infinity. Instead, a rescaled spectrum called the greybody spectrum is observed \cite{Singleton2011Aug,Akhmedova2008Aug,
Maldacena1997Jan,Cvetic1997Oct}. A measure of the deviation between the rescaled spectrum and the blackbody spectrum is given by the greybody factor. From an another perspective, greybody factor is actually a transmission coefficient in black hole scattering, where the curvature of spacetime plays the role of a potential barrier.
\begin{figure}[!htb]
\centerline{\includegraphics[scale=1]{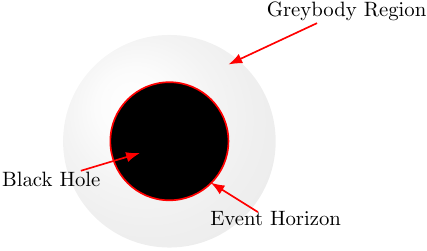}}
\caption{An illustration of greybody region around a BH is shown. Greybody factors, as one can see, are named for the way they modify the BH's spectrum, transforming it from that of a `black-body' to that of a `grey-body'.
}
\label{grey_illus}
\end{figure}

By inspection of the Schrödinger-like equation (cf. Eq. \eqref{schrod2}), the potential \(V(r)\) clearly acts as a barrier that filters the propagation of the scalar field in the black hole spacetime. In this context, part of the field passes through the potential barrier while the rest is reflected, which modulates the properties of Hawking radiation and any other fields moving through this curved background. Since the potential \(V(r)\) is determined only by the geometry of spacetime, its function in filtering the radiation directly stems from the gravitational structure. The greybody factor, which is qualitatively identical to the absorption cross section \(\sigma_{\text{abs}}\) of a black hole, indicates how much of the incoming field penetrates the barrier and is subsequently absorbed. In consequence, the radiation spectrum of the black hole is not exactly similar to that of a perfect black body; rather, it effectively resembles a grey body spectrum. This change in the spectral profile led to the term ``greybody factor," which emphasizes the interrelationship of quantum mechanical tunneling phenomenon to the intrinsic geometry of the BH spacetime. An illustration of the formation of a greybody region is shown in Fig. \ref{grey_illus}.

There are several methods to compute the greybody factor. One of such methods is to find its lower bound instead its exact value, which can be termed as \textit{greybody bound}. We specifically follow the method given by Visser and Boonserm \cite{Visser1999Jan,Boonserm2008Nov}, where the greybody bound is given by
\begin{equation} 
T_b \geq \operatorname{sech}^2\left(\frac{1}{2 \omega} \int_{r_h}^{\infty}|V| \frac{d r}{f(r)}\right)
\end{equation}
where $r_h$ is the horizon radius.
\begin{figure*}[htb]
\centerline{\includegraphics[scale=0.4]{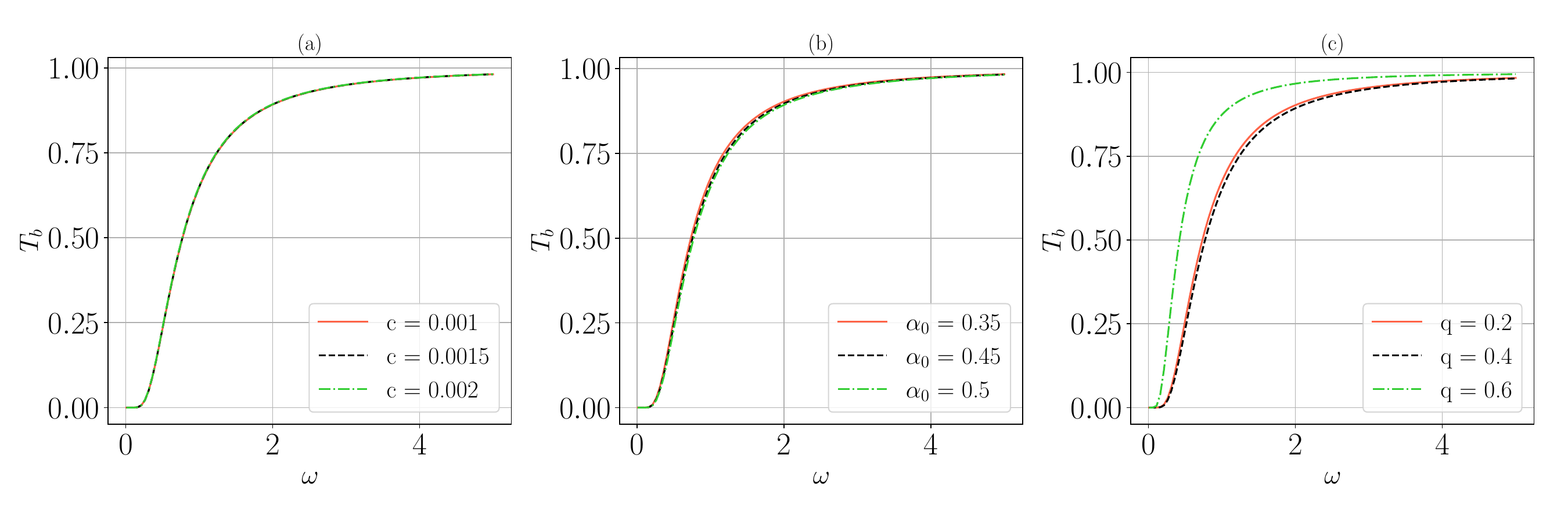}}
\caption{Greybody bounds for different choices of $c$, $\alpha_0$ and $q$ for $w = -2/3$ and $l=1$}
\label{greyl1_plot}
\end{figure*}
\begin{figure*}[htb]
\centerline{\includegraphics[scale=0.4]{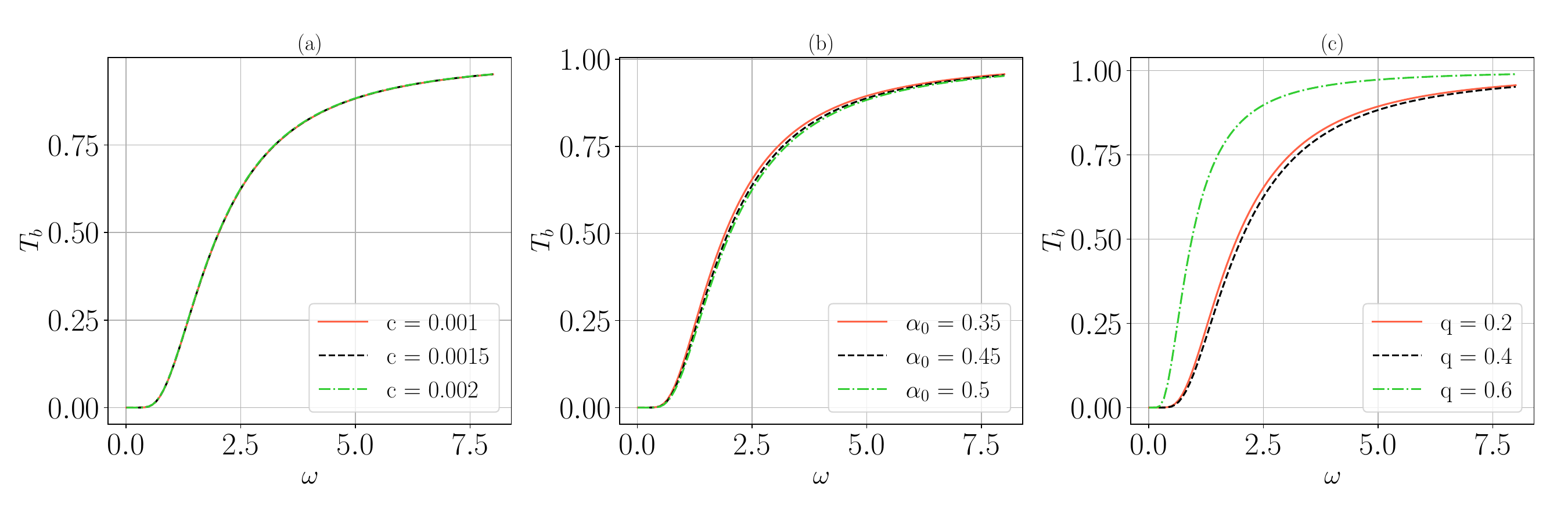}}
\caption{Greybody factor for different choices of $c$, $\alpha_0$ and $q$ for $w = -2/3$ and $l=2$}
\label{greyl2_plot}
\end{figure*}

To see how the greybody bounds depends on $l$, it is found that for lower multipole moment modes, $l = 1$, the transmission coefficient rises more quickly with frequency, suggesting that these modes experience a weaker effective potential barrier and thus have a higher chance of escaping. But, for higher values $l = 2$, the greybody factor consistently remains lower for all frequencies, which indicating that higher $l$ modes experience stronger centrifugal repulsion, which increases reflection and decreases transmission efficiency. This pattern arises from the modified Regge-Wheeler potential, where a rise in $l$ results in a greater height of the effective potential barrier, thus preventing the transmission of lower-energy waves.

The greybody factors of the Frolov BH is also influenced by additional BH parameters, such as the charge $q$, the Hubble length $\alpha_0$, and the quintessence parameter $c$. The presence of an electric charge modifies the effective potential, typically increasing its height and leading to stronger reflection, thereby reducing the transmission probability, particularly at intermediate frequencies. The quintessence field, parameterized by $c$, introduces an additional effect in the metric, which influences the effective potential in a manner dependent on the EoS of the surrounding field. A higher value of $c$ generally modifies the location and reduces the height of the barrier potential, thereby improving the transmission of the low-energy modes. This is one of the results in this study. Nevertheless, variations in the choice of the $c$ parameter in this work, does not drastically lead to observable shifts in the greybody profile. The Hubble length parameter $\alpha_0$, affects the potential barrier structure, thus modifying the transmission of the waves. As the value of $\alpha_0$ decreases, the height of the slope becomes steeper, which indicates an increased intensity of opposition to the transmission of the waves. The dependency on $\alpha$ is however found to be very weak, but noticeble for both $l = 1$ and $l = 2$ cases. 

Additionally, the behavior of $T_b$ in response to the charge q is seen to be quite prominent. That is, as the charge parameter increases, the slope of the curve also increases. This suggests that as q increases, the resistance to transmission of outgoing waves also rises. This behaviour is further amplified as $l$ is increased from 1 to 2.

\section{Weak Lensing}
\label{sec5}
\begin{figure*}[htb]
\centerline{\includegraphics[scale=0.8]{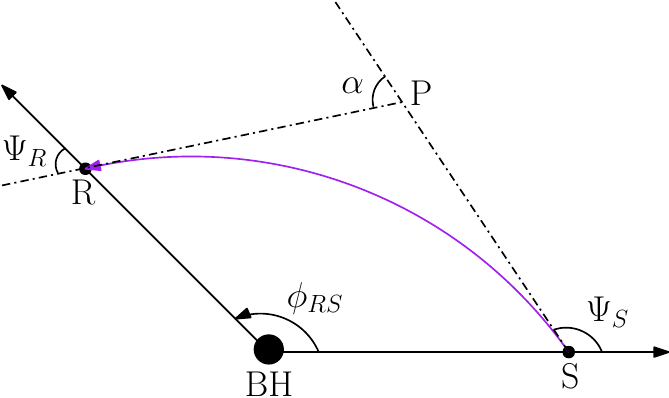}}
\caption{An illustration of the deflection angle of light trajectories around the BH is shown.}
\label{illus}
\end{figure*}

In this section, we shall calculate the weak deflection angle of light around the Frolov black hole, assuming the presence of a quintessence field. The intention of investigating the weak deflection angle is to figure out the effect of quintessence and other associated Frolov BH parameters on the trajectory of light beams.  In case of weak lensing, the light source and receiver are placed at finite radial distances \(r = r_{\text{S}} \) and \(r = r_{\text{R}} \) from the black hole, respectively (see Fig. \ref{illus}). In such a picture, the deflection angle can be defined as follows \cite{Ishihara2016Oct}: 
\begin{equation}
	\alpha \equiv \Psi_{\text{R}} - \Psi_{\text{S}} - \phi_{\text{RS}}
	\label{gravitational deflection angle}
\end{equation}
where \( \Psi_{\text{R}} \) and \( \Psi_{\text{S}} \) are the angles between the direction of light propagation and the radial direction at the receiver's position (\( R \)) and the source's position (\( S \)), respectively. The term \( \phi_{\text{RS}} \) denotes the change in the azimuthal angle \( \phi \) arising due to bending of light rays. Now, for a spherically symmetric spacetime, the metric in Eq. (\ref{line_el}), the angle between the direction of light propagation and the radial direction is determined by \cite{Ishihara2016Oct,Su2024Oct}:
\begin{equation}
	\sin \Psi = \frac{b \cdot \sqrt{f(r)}}{r} 
	\label{angle psi}
\end{equation}
Under the thin lens approximation, this definition of the gravitational deflection angle, applicable for finite source and observer distances, is illustrated in Fig. \ref{illus}. In the figure, $S$ and $R$ denotes the coordinates of the source and receiver and $\Psi_R$ and $\Psi_S$ are the angles between the direction of propagation of light and the radial directions at the receiver’s position ($R$)
and the source’s position ($S$), respectively. In the thin lens approximation, the deflection angle \( \alpha \equiv \Psi_{\text{R}} - \Psi_{\text{S}} - \Delta \phi_{\text{RS}} \) can be demonstrated through the quadrilateral. The exterior angles at each vertex of the quadrilateral are \( \Psi_{\text{S}} \), \( \alpha \), \( \pi - \Psi_{\text{O}} \), and \( \pi - \Delta \phi_{\text{RS}} \), respectively. In this approximation one can assume that the spacetime is flat everywhere except at the thin lens position (the central black hole). As a result, the sum of the exterior angles equals \( 2\pi \), which explains the definition of the deflection angle \( \alpha \equiv \Psi_{\text{R}} - \Psi_{\text{S}} - \Delta \phi_{\text{RS}} \).

In spherically symmetric spacetimes, the angle $\phi_{RS}$ can be derived from the geodesic equation 
\begin{equation}
	\frac{dr}{d\phi} = \frac{dr}{d\lambda} \cdot \frac{d\lambda}{d\phi}
	= \pm r^{2} \sqrt{\frac{1}{b^{2}}-\frac{f(r)}{r^{2}}}
\end{equation}
where $\lambda$ is the affine parameter, and $\epsilon=0$ for null rays. It is straightforward that the conserved quantities are the energy $E = f(r) \frac{dt}{d\lambda} = f(r) \dot{t}$, and the angular momentum $L = r^2 \sin^{2}\theta \frac{d\phi}{d\lambda} = r^2 \sin^{2}\theta \dot{\phi} $, which can be used to define the impact parameter through $b = |L/E|$.
\begin{widetext}
\begin{subequations}
	\begin{eqnarray}
		&& \left.\frac{dr}{d\phi} = - r^{2} \sqrt{\frac{1}{b^{2}}-\frac{f(r)}{r^{2}}} < 0 \quad  \right\} \ \ \text{photon going from S($r=r_{\text{S}}$) to the closest approach point $r=r_{0}$} 
		\nonumber
		\\
		&& \left. \frac{dr}{d\phi} = r^{2} \sqrt{\frac{1}{b^{2}}-\frac{f(r)}{r^{2}}} > 0 \quad  \right\} \ \ \text{photon going from the closest approach point $r=r_{0}$ to  R($r=r_{\text{R}}$})
		\nonumber
	\end{eqnarray}
\end{subequations} 
\end{widetext}

Now using the angle $\Psi$ from Eq. (\ref{angle psi}) and the angle $\phi_{RS}$, the weak deflection angle of can be expressed in a general form as \cite{Su2024Oct}
\begin{widetext}
\begin{eqnarray}
	\alpha = \Psi_{\text{R}} - \Psi_{\text{S}} - \Delta \phi_{\text{RS}}
	& = & \Psi_{\text{R}} - \Psi_{\text{S}} 
	+ \int_{r_{\text{S}}}^{r_{0}} \frac{d\phi}{dr} dr 
	+ \int_{r_{0}}^{r_{\text{R}}} \frac{d\phi}{dr} dr \nonumber
	\\
	& = & \arcsin \bigg( \frac{b \sqrt{f(r_{\text{R}})}}{r_{\text{R}}} \bigg)
	+ \arcsin \bigg( \frac{b \sqrt{f(r_{\text{S}})}}{r_{\text{S}}} \bigg) - \pi
	+ \int_{r_{0}}^{r_{\text{S}}} \frac{dr}{r^{2}\sqrt{\frac{1}{b^{2}}-\frac{f(r)}{r^{2}}}}
	+ \int_{r_{0}}^{r_{\text{R}}} \frac{dr}{r^{2}\sqrt{\frac{1}{b^{2}}-\frac{f(r)}{r^{2}}}} \nonumber
	\\
	\label{gravdef}
\end{eqnarray}
\end{widetext}
Like shown in Fig. \ref{illus}, the angle $\Psi$ at the coordinate of light source $S$ is an obtuse angle given by $\Psi_{\text{S}} = \pi - \arcsin \big( \frac{b \sqrt{f(r_{\text{S}})}}{r_{\text{S}}} \big)$.

\begin{figure}[htb]
\centerline{\includegraphics[scale=0.4]{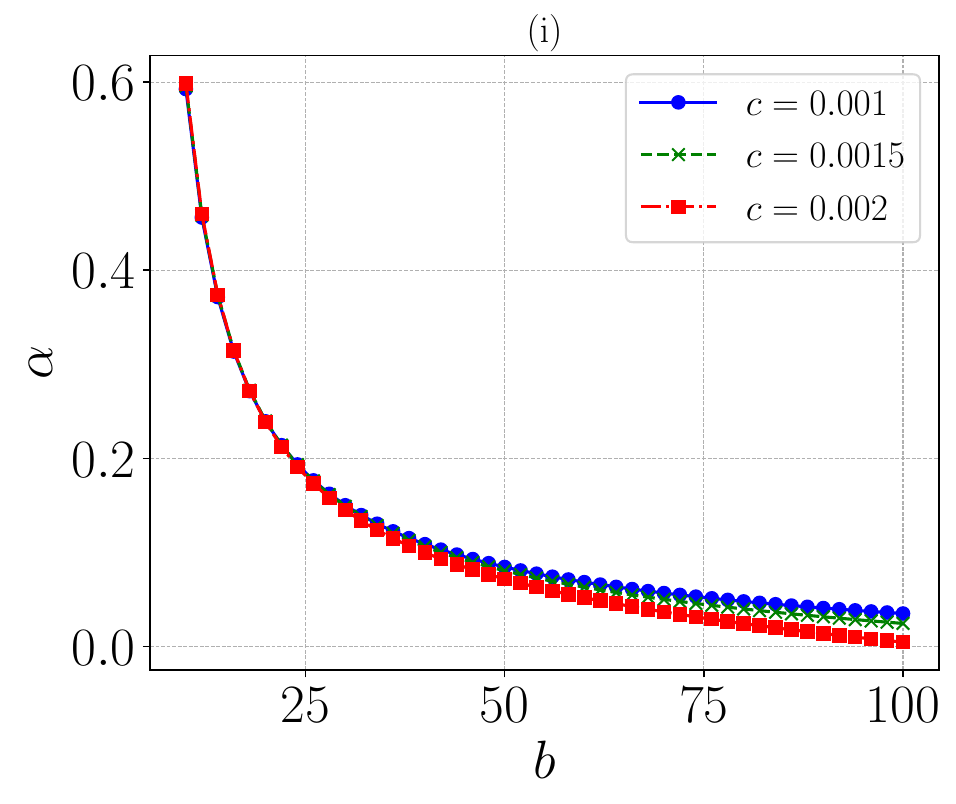}}
\centerline{\includegraphics[scale=0.4]{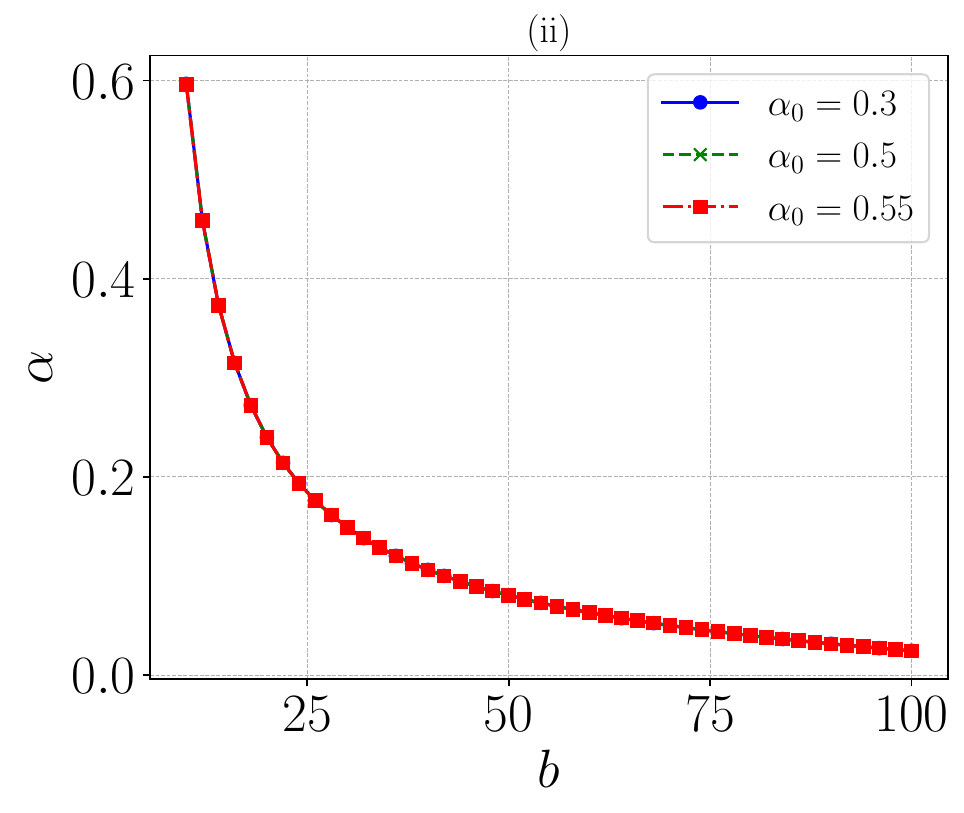}} 
\centerline{\includegraphics[scale=0.4]{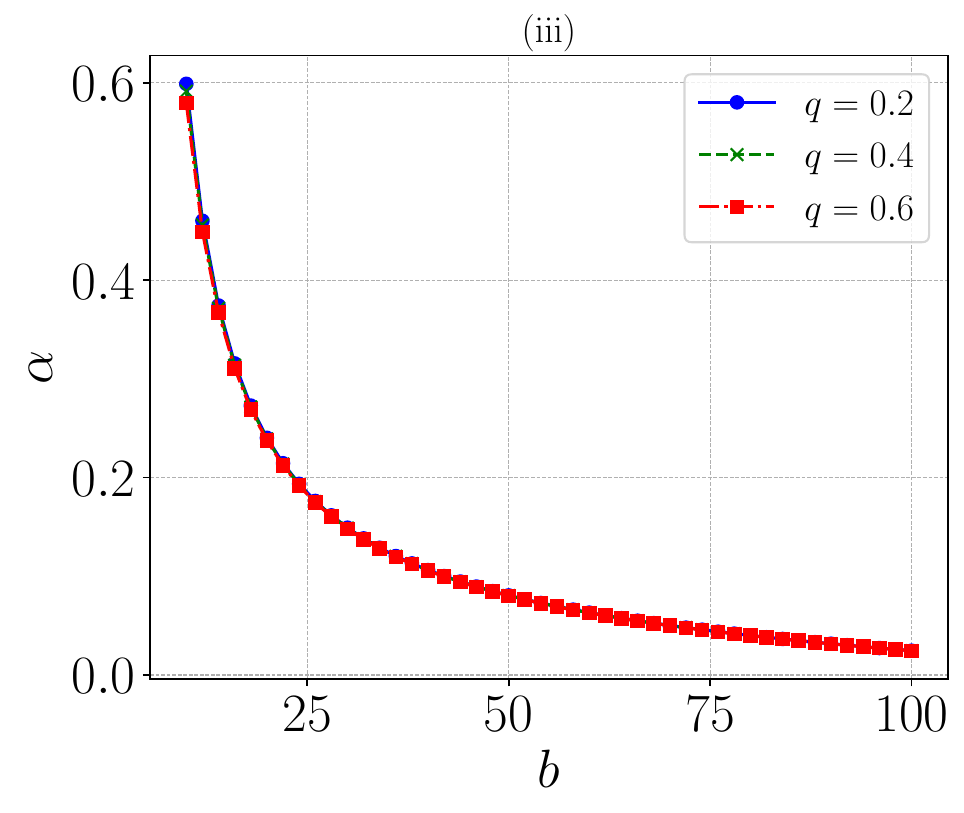}}
\caption{The weak deflection angle is plotted against the impact parameter for three values of $c$, $\alpha_0$ and $q$. In the plots (i), (ii) and (iii), $(\alpha_0, q) = (0.5, 0.4)$,$(c, q) = (0.0015, 0.4)$ and $(\alpha_0, c) = (0.5, 0.0015)$ are kept fixed respectively. }
\label{lensing1}
\end{figure}

With the help of Eq. \eqref{gravdef}, the deflection angle $\alpha$ is solved numerically and plotted against the impact parameter $b$ for different combinations of the model parameters $c$, $\alpha_0$, and $q$. The results are shown in Fig. \ref{lensing1}(i)--(iii). As seen from Fig. \eqref{lensing1}(i), the deflection angle decreases monotonically with a larger impact parameter $b$. For this case, we choose three different values of $c$, whereas $\alpha_0$ and $q$ are fixed at $0.5$ and $0.4$, respectively. As can be observed, a mild variation in the deflection angle is apparent for larger values of $b$, with greater $c$ values corresponding to an increased rate of decrease in $\alpha$. A higher $c$ parameter,  associated with the density or spatial distribution of quintessence, might enhance its repulsive gravitational influence at larger distances and thus decrease light-bending even more as $b$ increases. Contrastingly, in Fig. \ref{lensing1} (ii) and (ii), where $(c, q) = (0.0015, 0.4)$ and $(c, \alpha_0) = (0.0015, 0.5)$ are kept fixed respectively, the variation in the deflection angle is quite insensitive to the parameters $\alpha_0$ and $q$, as no noticeable variation in $\alpha$ is seen. 

%\begin{figure}[htb]
%\centerline{\includegraphics[scale=0.36]{lensing_c_w-2by3.pdf}\hspace{0.1cm} \includegraphics[scale=0.35]{lensing_alpha_w-2by3} \hspace{0.1cm} \includegraphics[scale=0.35]{lensing_q_w-2by3}}
%\caption{The weak deflection angle is plotted against the impact parameter for three values of $c$, $\alpha_0$ and $q$. In the plots (i), (ii) and (iii), $(\alpha_0, q) = (0.5, 0.4)$,$(c, q) = (0.0015, 0.4)$ and $(\alpha_0, c) = (0.5, 0.0015)$ are kept fixed respectively. }
%\label{lensing1}
%\end{figure}

\section{Chaotic behaviour: Lyapunov exponents}
\label{sec6}
The Lyapunov exponent is the average rate at which two neighbouring geodesics separation within phase plane. If the Lyapunov exponent is positive, these geodesics are said to diverge, while if it is negative, then they are said to converge. The fact that BH spacetimes have unstable circular geodesics also shows us that GR is non-linear, which reflects on a positive Lyapunov exponent. This non-linearity suggests that the system is not integrable but the circular geodesics can behave chaotically as a consequence.

The Lyapunov exponent ($\tilde{\lambda}$) is directly related to the effective potential through the relation \cite{Kumara2024Jan}
\begin{equation}
\tilde{\lambda}^2 = -\frac{(V_{eff})''}{2\dot{t}^2},
\label{lyap}
\end{equation}
where $\dot{t} = E f(r)^{-1}$ and $V_{eff} = f(r) \left(\frac{E^2}{f(r)}+\frac{L^2}{2 r^2}\right)$ are the derivative of the coordinate $t$ with respect to a affine parameter (say $\mu$) and $V_{eff}$ is the effective potential for massless particles respectively.
Also, ($^{\prime \prime}$) represents the second derivative with respect to $r$. In the subsequent analysis, we can safely set $E = 1$ for simplicity. The value of $E$ is only responsible for the height of the peak of the effective potential, and does not affect the radius of the circular orbits. 

The circular geodesics are said to be unstable, stable and marginally stable if the Lyapunov exponent $\tilde{\lambda}$ attains real, imaginary and zero values. In our model, we calculate the Lyapunov exponent as
%\begin{widetext}
\begin{equation}
\begin{aligned}
\tilde{\lambda}^2 &= 
\frac{1}{4 E^2 
\left(2 \alpha_0^2 M r_c + r_c^4 + \alpha_0^2 q^2\right)^5} \\
&\quad \times \left[ 
L^2 r_c^{-9w-7} 
\left(c \left(2 \alpha_0^2 M r_c + r_c^4 + \alpha_0^2 q^2\right) \right. \right. \\
&\qquad\quad \left. - r_c^{3w+1} \left(2 \alpha_0^2 M r_c - 2 M r_c^3 
+ q^2 \left(\alpha_0^2 + r_c^2\right) + r_c^4\right)\right)^2 \\
&\quad \times \Bigg( 3 c \left(3 w^2 + 7 w + 4\right) 
\left(2 \alpha_0^2 M r_c + r_c^4 + \alpha_0^2 q^2\right)^3 \\
&\qquad - 2 r_c^{3w+1} \Big( 
q^2 r_c^2 \left(8 \alpha_0^4 M^2 r_c^2 + 24 \alpha_0^2 M r_c^5 
+ 36 \alpha_0^4 M r_c^3 \right. \\
&\qquad\quad \left. + 9 \alpha_0^2 r_c^6  + 10 r_c^8 + 36 \alpha_0^6 M^2 \right) \\
&\qquad\quad + 3 r_c^3 \left(4 \alpha_0^2 M^2 r_c^3 
\left(3 \alpha_0^2 + r_c^2\right) + M \left(6 \alpha_0^2 r_c^6 
- 4 r_c^8\right) \right. \\
&\qquad\quad \left. + r_c^9 + 8 \alpha_0^6 M^3 \right) \\
&\qquad\quad + q^4 \left(18 \alpha_0^6 M r_c 
- 6 \alpha_0^2 r_c^6 + 9 \alpha_0^4 r_c^4\right) 
+ 3 \alpha_0^6 q^6 \Big) \Bigg) 
\Bigg].
\end{aligned}
\label{lyap2}
\end{equation}
%\end{widetext}

\begin{figure*}[htb]
\centerline{\includegraphics[scale=0.45]{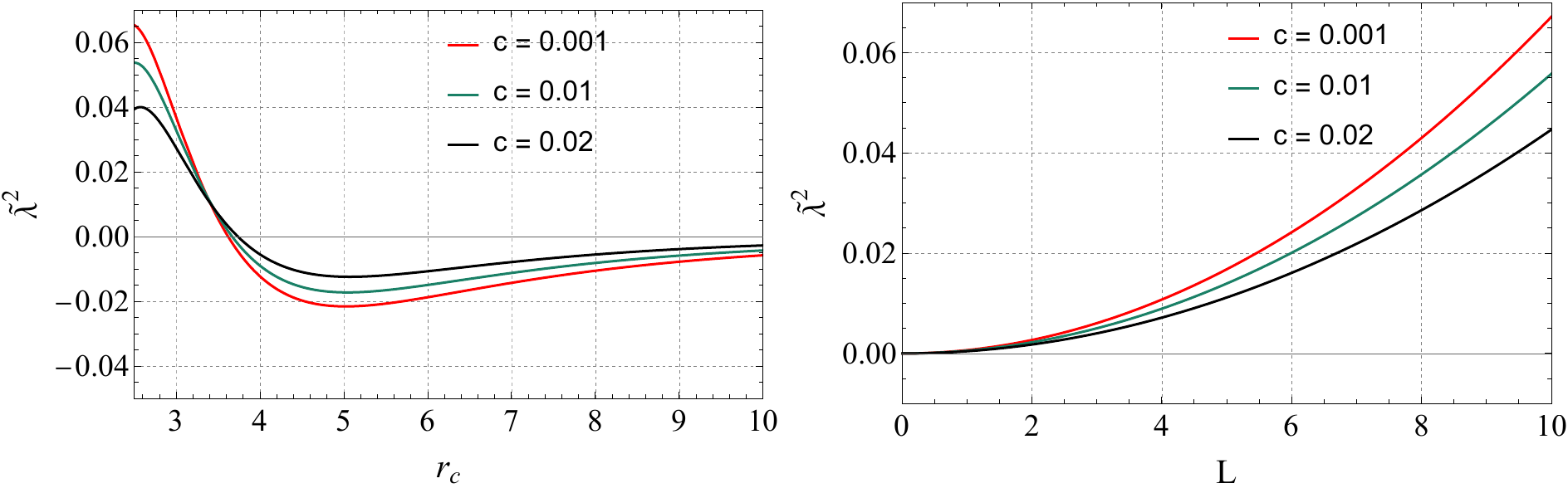}}
\centerline{\includegraphics[scale=0.45]{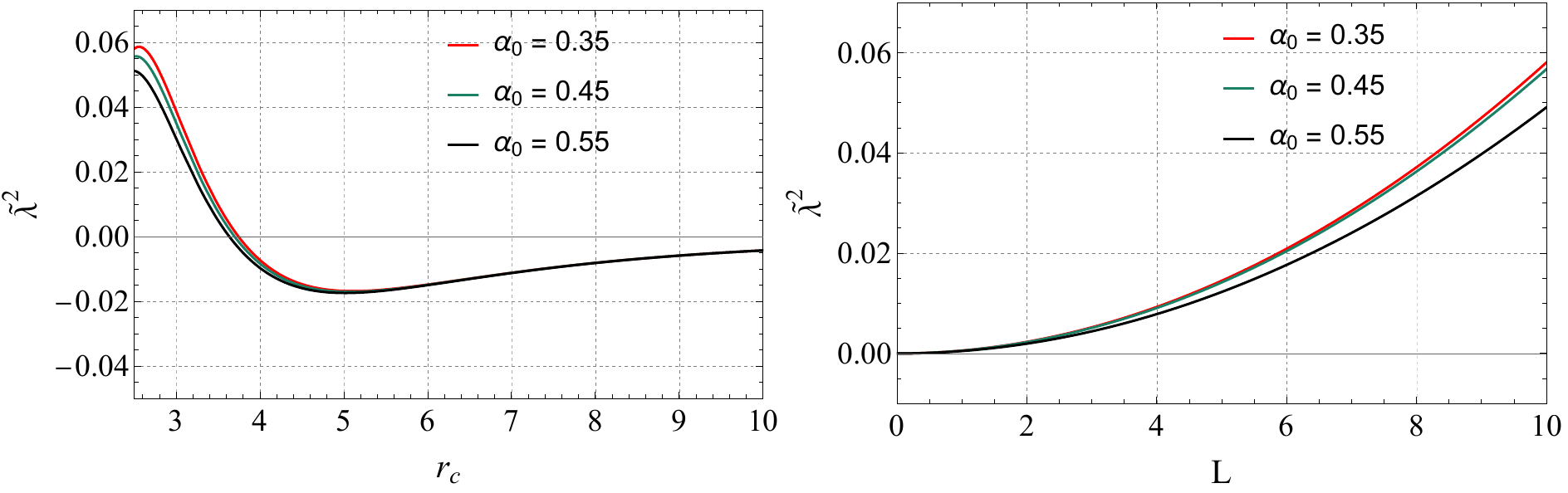}}
\centerline{\includegraphics[scale=0.45]{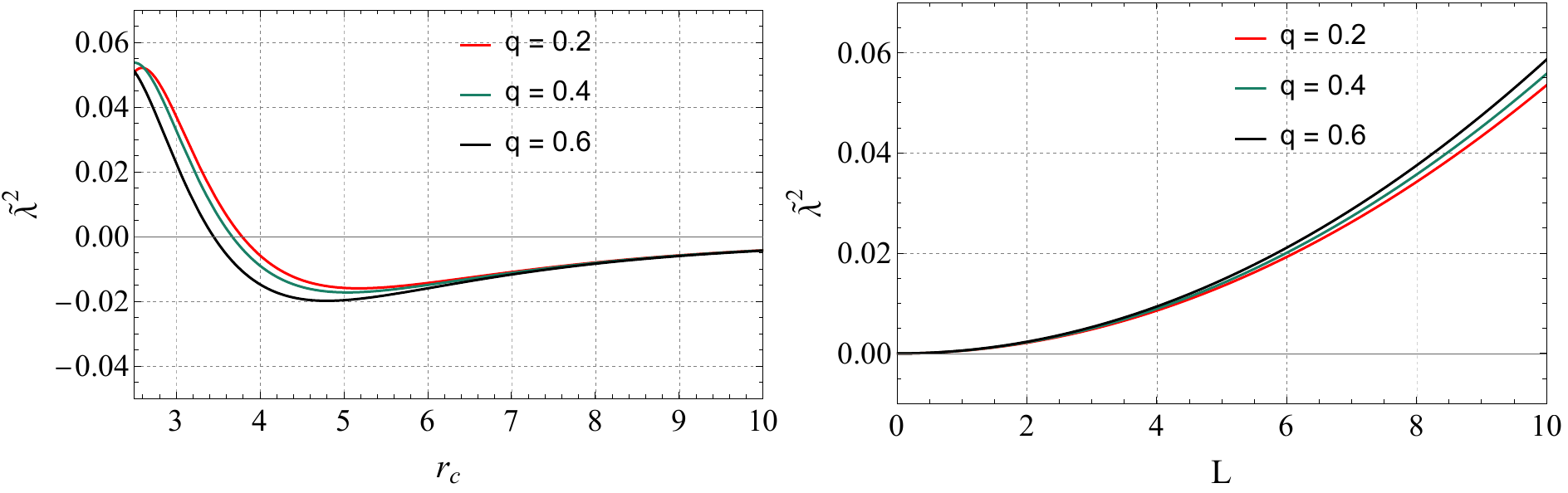}}
\caption{The Lyapunov exponent $\tilde{\lambda^2}$ is plotted with respect to $r_c$ and $L$ for different values of $c$ (top panel), $\alpha_0$ (middle panel) and $q$ (bottom panel).}
\label{lyap_main}
\end{figure*}

The behavior of the squared Lyapunov exponent \( \lambda^2 \) as a function of circular orbit radius \( r_c \) is dictated by the requirement \( \dot{r}=0 \). In Fig.~\ref{lyap_main}, one can see that for smaller radii of \( r_c \), \( \lambda^2 \) is positive, showing that the circular photon orbits are unstable because the nearby trajectories experience exponential divergence. For large values of radii, however, \( \lambda^2 \) becomes negative, which marks a marginally stable regime. Additionally, the figures suggest that as angular momentum \(L\) is increased, \(\lambda^2\) also increases monotonically, that emphasizes the correspondence of orbital instability with angular momentum within the black hole spacetimes.

Here, the vertical axis measures the square Lyapunov exponent \(\lambda^2\), quantitatively computing the stability of circular orbits of photons for black holes. All the panels show \(\lambda^2\) vs. either the orbit radius \(r_c\) (left column) or the angular momentum of the photon \(L\) (right column) with the quintessence parameter \(c\), the length scale \(\alpha_0\), and the black hole charge \(q\) varied systematically. A positive values of \(\tilde{\lambda}^2\) means that small perturbations in the trajectory of the photon will grow exponentially, which a signifies dynamical instability, while a negative value of $\tilde{\lambda}^2$ signifies a stable or marginally stable photon orbit.

To predict the existence of chaotic behaviour, one can calculate the chaos bound of the null circular geodesics. Chaos is an interesting phenomenon observed in non-linear physical systems, which represents unpredictable random motion in some deterministic but non-linear systems sourcing from the susceptibility to the initial conditions. It is known for a general chaotic system, that if chaos is mathematically expressed as a function of time $C(t)$, it grows exponentially over time, i.e. \( C(t) \approx \exp(\tilde{\lambda} t) \), where \( \tilde{\lambda} \) is called the Lyapunov exponent, which manifests as the system's sensitivity to the initial conditions. Maldacena, Shenker and Stanford had proposed a conjecture that there should exist a universal upper bound for the Lyapunov exponent in general BH thermodynamic systems \cite{Maldacena2016Aug}, i.e. 
\begin{equation}
\tilde{\lambda} \le \frac{\kappa}{\hbar},
\label{chaos_bound}
\end{equation}
where $\kappa$ represents the surface gravity of the BH which is associated with the temperature of the BH through $T=\kappa/2\pi$. This is known as the chaos bound. 
Since, the BH surface gravity $\kappa$ is defined as $\kappa = f'(r)/2$, it is straight forward to calculate $\tilde{\lambda}^2 - \kappa^2$ and considering units where $\hbar = 1$, this gives
\begin{widetext}
\begin{equation}
\begin{aligned}
\tilde{\lambda}^2 - \kappa^2 &= 
\frac{1}{4 \big(2 \alpha_0^2 M r_c + r_c^4 + \alpha_0^2 q^2\big)^5} \times \Bigg[ \\
& \quad r_c^{-9 w - 7} \Bigg( 
\frac{L^2 \Big(c \big(2 \alpha_0^2 M r_c + r_c^4 + \alpha_0^2 q^2\big) 
- r_c^{3 w+1} \big(2 \alpha_0^2 M r_c - 2 M r_c^3 
+ q^2 \big(\alpha_0^2 + r_c^2\big) + r_c^4\big)\Big)^2}{E^2} \Bigg) \\
& \quad \times \Bigg( 
3 c \big(3 w^2 + 7 w + 4\big) 
\big(2 \alpha_0^2 M r_c + r_c^4 + \alpha_0^2 q^2\big)^3 
- 2 r_c^{3 w+1} \Bigg(q^2 r_c^2 \Big(8 \alpha_0^4 M^2 r_c^2 
+ 24 \alpha_0^2 M r_c^5 + 36 \alpha_0^4 M r_c^3 \\
& \quad + 9 \alpha_0^2 r_c^6 + 10 r_c^8 
+ 36 \alpha_0^6 M^2\Big) 
+ 3 r_c^3 \Big(4 \alpha_0^2 M^2 r_c^3 \big(3 \alpha_0^2 + r_c^2\big) 
+ M \big(6 \alpha_0^2 r_c^6 - 4 r_c^8\big) 
+ r_c^9 + 8 \alpha_0^6 M^3\Big) \\
& \quad + q^4 \Big(18 \alpha_0^6 M r_c - 6 \alpha_0^2 r_c^6 
+ 9 \alpha_0^4 r_c^4\Big) + 3 \alpha_0^6 q^6\Bigg)\Bigg) \\
& \quad - r_c^{3 w+3} \big(2 \alpha_0^2 M r_c + r_c^4 
+ \alpha_0^2 q^2\big) \bigg(c (3 w+1) 
\big(2 \alpha_0^2 M r_c + r_c^4 + \alpha_0^2 q^2\big)^2 \\
& \quad + 2 r_c^{3 w+3} \Big(-q^2 \big(2 \alpha_0^2 M r_c + r_c^4\big) 
+ M r_c^2 \big(r_c^3 - 4 \alpha_0^2 M\big) 
+ \alpha_0^2 q^4\Big)\bigg)^2\Bigg] .
\end{aligned}
\label{ch_bound}
\end{equation}
\end{widetext}
\begin{figure*}[tbh]
\centerline{\includegraphics[scale=0.45]{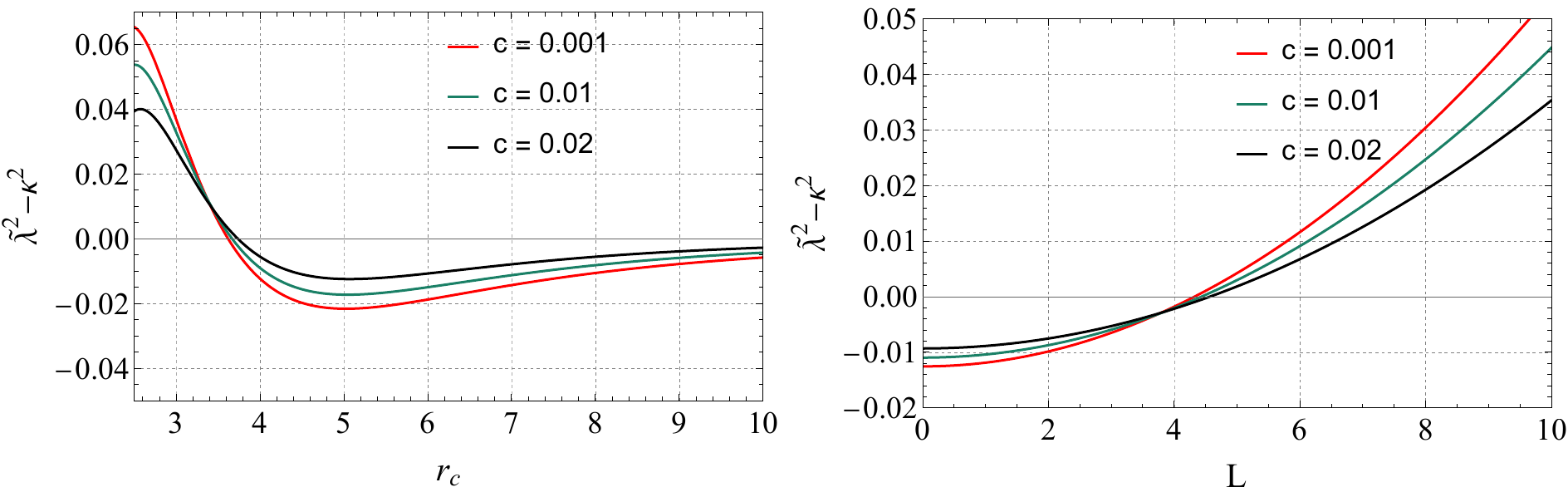}}
\centerline{\includegraphics[scale=0.45]{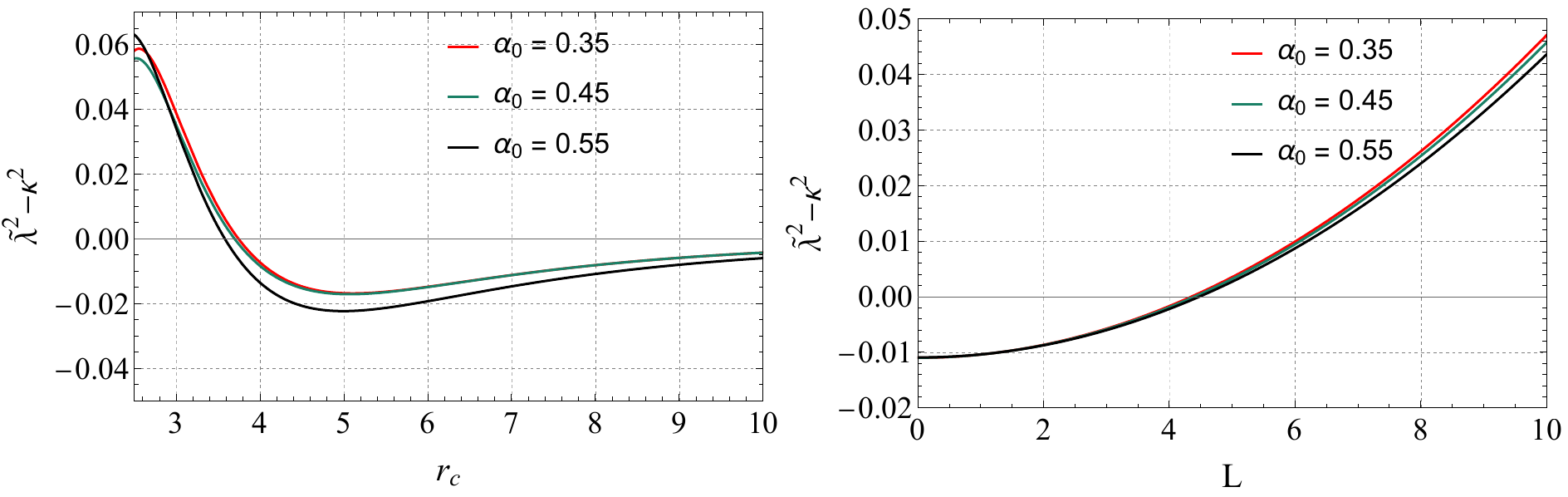}}
\centerline{\includegraphics[scale=0.45]{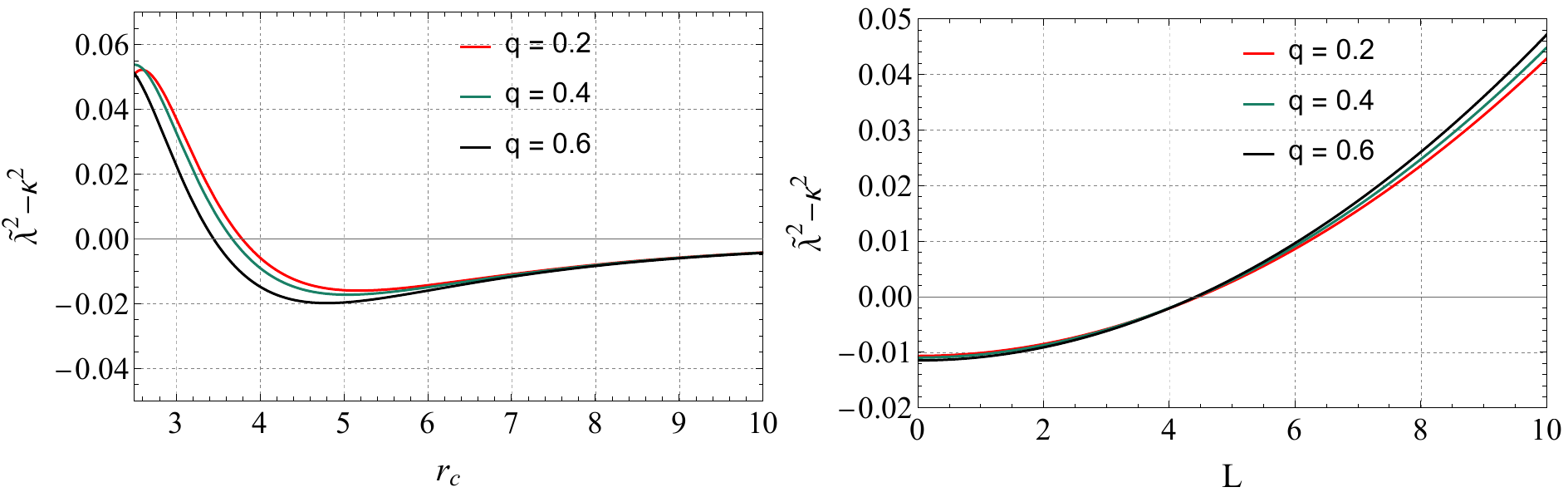}}
%\centerline{(a) $w = -2/3$}
%\vspace{0.5cm}
%\centerline{\includegraphics[scale=0.5]{chaosbound_new2.pdf}}
%\centerline{(b) $w = -4/9$}
\caption{The plot of chaos bound is shown for $w = -2/3$ for different values of $c$ (top panel), $\alpha_0$ (middle panel) and $q$ (bottom panel).}
\label{chaos_bound_fig}
\end{figure*}
Next, in Fig. \ref{chaos_bound_fig}, we have plotted the behaviour of \( \tilde{\lambda}^2 - \kappa^2 \) with respect to both the photon orbit radius $r_c$ and the angular momentum \( L \). Since, by definition, the chaos bound holds when \( \tilde{\lambda}^2 - \kappa^2 < 0 \) and is violated when \( \tilde{\lambda}^2 - \kappa^2 > 0 \), we find that for larger photon orbit radii the chaos bound is satisfied, and at smaller photon orbit radii it is violated, which implies that chaotic motion is more likely to be found in the near-horizon regime of the BH. Also, the bound holds at lower values of angular momentum and violates at higher values. This indicates that higher angular momentum photons tend to have chaotic orbits. Furthermore, the black hole parameters \(c\) , \(\alpha_0\) and \(q\)  play important roles in the chaos bound. An increase in the value of \(c\)  shifts the radius at which the bound is broken towards the horizon. The parameter \(\alpha_0\) and \(q\) also acts in a similar way shifting the critical radius where the bound is violated nearer to the horizon as they increase. In contrast, the critical value of angular momentum where the chaos bound is violated is very feebly affected by $c$ values whereas $\alpha_0$ and $q$ does not practically affect the critical value of bound violation.

\subsection{Scattering Cross Section}
We finally determine the high-energy absorption cross-section through the Sinc approximation. Initially, Sanchez investigated the absorption cross-section for the Schwarzschild black hole in the UV regime and suggested that, as the frequency increases for a typical material sphere, the absorption cross-section grows monotonically while oscillating around the constant geometric-optics value associated with the black hole’s photon sphere \cite{Sanchez1978Aug}. Next, the relationship between the impact parameter and the photon sphere’s cross-section is established at critical values, thereby constraining the absorption cross-section. It follows that, at low energy scales, the absorption cross-section reflects the intrinsic properties of the black hole, matching its area, as shown by Das et al \cite{Das1997Jan}. In contrast, at high energies, the geometric cross-section of the photon sphere can be examined using complex angular momentum methods \cite{Decanini2011Feb}, where Decanini et al. employed the Regge pole techniques to confirm that the high-energy absorption cross-section exhibits an oscillatory behaviour described by a $Sinc(x)$ function involving the photon sphere.

\begin{figure*}[tbh]
\centerline{\includegraphics[scale=0.45]{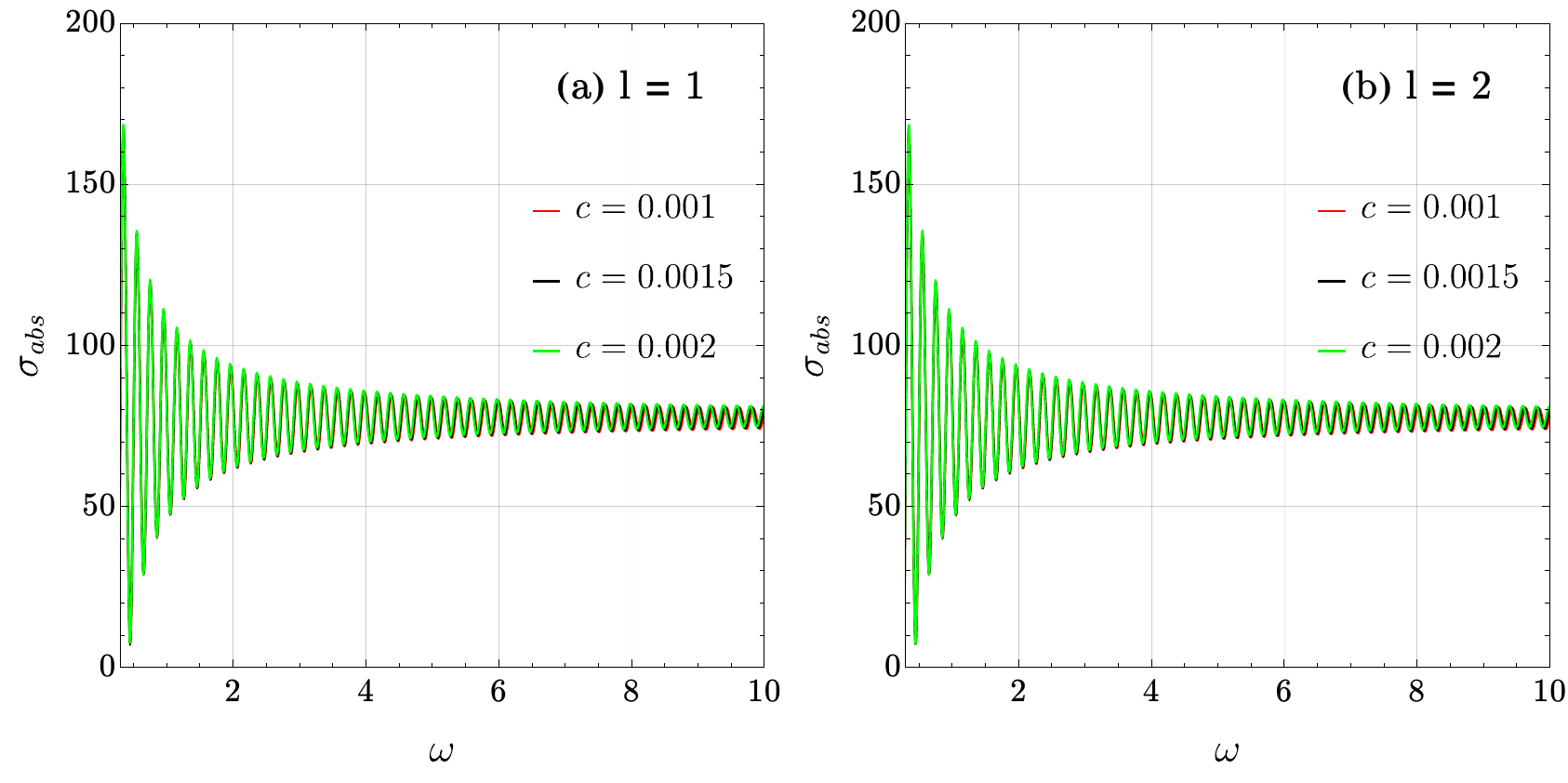}}
\centerline{\includegraphics[scale=0.45]{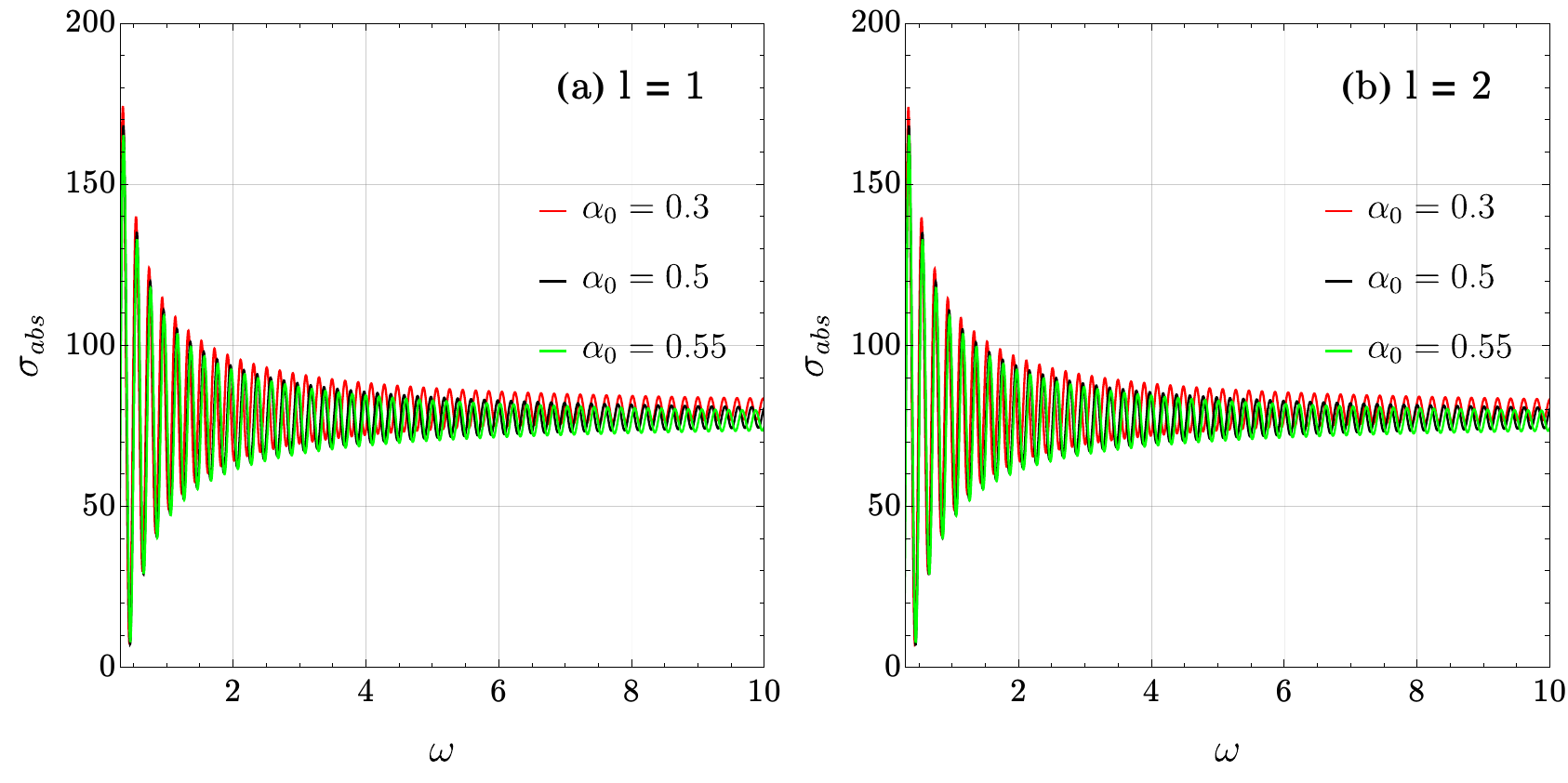}}
\centerline{\includegraphics[scale=0.45]{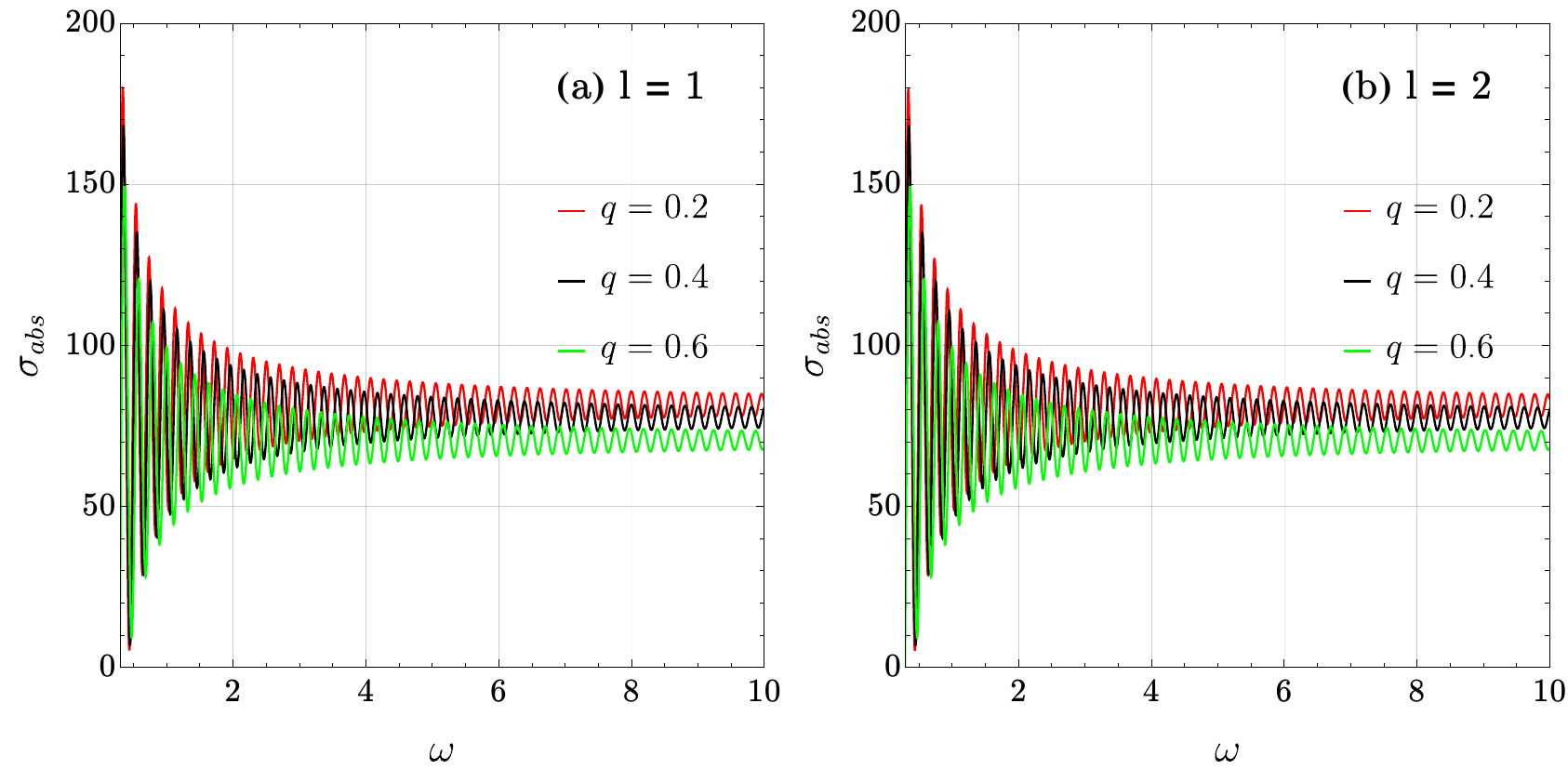}}
\caption{The total absorption cross section is shown for different values of the model parameters and for different $l$ values of the effective potential. }
\label{sig_plot}
\end{figure*}

To calculate the total absorption cross section, one may use the Sinc approximation, which states that the total absorption cross section at the eikonal limit is given by $\sigma_{abs} \approx \sigma_{geo} + \sigma_{osc}$ \cite{Decanini2011Feb},
where $\sigma_{geo}$ is the classical capture cross section of photon geodesics, which reduces from the total absorption cross section at the high frequency limit. It is related to the critical impact parameter $b_{crit}$ through $\sigma_{geo} = \pi b^2_{crit}$. Also, the oscillatory counterpart of the total cross section $\sigma_{osc}$ at the eikonal limit can be given by:
\begin{equation}
\sigma_{osc} = - \frac{4\pi \tilde{\lambda} b^2_{crit}}{\omega} \exp \left(-\pi \tilde{\lambda}\right) \sin \frac{2\pi\omega}{\Omega_{ph}},
\label{sig_osc}
\end{equation}
where $\tilde{\lambda}$ is the Lyapunov exponent, $\Omega_{ph}$ is the angular velocity at the critical orbit and $r_{ph}$ is the radius of the photon sphere.

The total absorption cross section is plotted and shown in Fig. \ref{sig_plot} (top panel). We can see that there is no drastic variation for different $c$ values (when $\alpha_0$ and $q$ are kept fixed at 0.5 and 0.4 respectively) at the low frequency limit, but a slight variation is seen at high frequencies. Nevertheless, the variation is very negligible. Also, no variation is seen as $l$ increases from 1 to 2. Next, concerning the values of $\alpha_0$ parameter, shown in Fig. \ref{sig_plot} (middle panel), the total absorption cross section increases with decreasing values of $\alpha_0$ (when $c$ and $q$ are fixed at 0.0015 and 0.4 respectively). But, no effect is seen when $l$ is increased from $1$ to $2$. Finally, the total absorption cross section is strongly influenced by the charge parameter $q$, where there is a visible effect on the amplitude of the total absorption cross section, as shown in Fig. \ref{sig_plot} (bottom panel) It is seen that as the charge parameter decreases, the absorption cross section increases, irrespective of $l$ values. In all the cases, however, a regular oscillation is maintained at the high frequency regime.

\section{Conclusion}
\label{conc}
In the first part, we analyzed the behavior of scalar quasinormal frequencies within the context of a Frolov BH in a quintessence field. We studied the quasinormal modes of the BH system by employing the first order WKB approximation to analyzed how the Frolov BH coupled with the quintessence field dynamically responds to scalar perturbations. We have found that, with increasing values of $c$, the real part of the QNM frequency, drops, which corresponds to a decrease of the oscillatory behaviour of the spacetime as the intensity of the quintessence field increases. At the same time, an occurrence of less negative values of the QNM frequency in the imaginary counterpart, implying perturbations falling off more slowly. This suggests the effect of quintessence on the BH system's dynamical response against scalar perturbations.

We have also analyzed the dependence of greybody factors on the multipole moment $l$ and black hole parameters $q$, $\alpha_0$, and $c$. Lower multipole modes ($l = 1$) experience weaker barriers, enhancing transmission, while higher modes ($l = 2$) face stronger repulsion, reducing transmission efficiency. The charge $q$ increases the barrier height, suppressing transmission, especially for higher $l$. The quintessence parameter $c$ affects low-energy modes, while smaller $\alpha_0$ steepens the barrier, reducing transmission. The transmission coefficient $T_b$ shows strong sensitivity to $q$, further amplified for higher $l$, providing insights into wave propagation in modified black hole spacetimes.

We have studied the consequences of weak lensing around the BH system. The deflection angle of weak lensing $\alpha$ decreases monotonically with increasing impact parameter $b$, with a greater rate of decrease for larger $c$, indicating an enhanced repulsive effect from quintessence. In contrast, variations in $\alpha_0$ and $q$ have negligible influence on $\alpha$, as observed in the nearly unchanged deflection profiles. This suggests that the quintessence parameter $c$ plays a dominant role in modifying light bending, while $\alpha_0$ and $q$ have minimal impact. The results highlight the significance of quintessence density in gravitational lensing effects around the black hole.

Finally, we analysed the chaos bound within the context of our BH systems. We find that, the chaos bound is satisfied for larger and violated for smaller radii of the circular photon orbits. This would in turn, imply that chaotic behaviour of photons in circular orbits is likely to manifest in the strong gravitational field nearer the BH. Moreover, the bound is obeyed for low angular momentum photons, while it is violated for high angular momentum photons. This suggests that high-angular-momentum photons are more likely to exhibit chaotic trajectories. Physically, the violation of the chaos bound near the BH indicates that the quintessence field indeed has strong influence on its thermodynamic properties (as chaos bound is directly related to surface gravity, and hence temperature of the BH). These results suggest that some modifications may be feasible in the traditional framework of BH thermodynamics based on Bekenstein-Hawking entropy under the influence of quintessence. Then making use of the Lyapunov exponents, we also calculated the total absorbtion cross section of radiation, which is intimately related to the critical photon orbit and impact parameter, which is found to be dependent mostly on the charge parameter and the Hubble length scale parameter, and depend minimally on the quintessence parameter. The quintessence field however, seems to play a key role in the primary thermodynamic properties such as the BH temperature, possibly leading to potential deviations from standard GR predictions in the strong-field regime.

\bibliography{bibliography.bib}
\end{document}